\documentclass[acmsmall,screen,nonacm]{acmart}
\usepackage{amsmath} 
\usepackage{graphicx}
\usepackage{booktabs,tabularx}
\usepackage{float}
\usepackage{xcolor}
\usepackage{soul}
\usepackage{colortbl}
\usepackage[breakable]{tcolorbox}
\usepackage{caption}
\usepackage{subcaption}
\DeclareCaptionLabelFormat{rq1}{#1 #2 (RQ1)}
\DeclareCaptionLabelFormat{rq2}{#1 #2 (RQ2)}
\DeclareCaptionLabelFormat{rq3a}{#1 #2 (RQ3-a)}
\DeclareCaptionLabelFormat{rq3b}{#1 #2 (RQ3-b)}
\DeclareCaptionLabelFormat{rq3c}{#1 #2 (RQ3-c)}
\usepackage{algorithm}
\usepackage{algpseudocode}

\graphicspath{{}}

\definecolor{lightgray}{gray}{0.95}

\setcopyright{acmlicensed}
\acmDOI{}

\usepackage{xspace}

\makeatletter
\DeclareRobustCommand\onedot{\futurelet\@let@token\@onedot}
\def\@onedot{\ifx\@let@token.\else.\null\fi\xspace}
\def\eg{{e.g}\onedot} 

\def\ie{{i.e}\onedot}

\def\etal{{et al}\onedot}

\newcommand{\rqone}{How does our method perform compared to existing testing approaches?\xspace}
\newcommand{\rqtwo}{How does the failure-detection efficiency vary with the testing-budget ratio?\xspace}
\newcommand{\rqthree}{To what extent do the fairness-aware modules contribute to the overall effectiveness?\xspace}

\newcommand{\method}{FairTest\xspace}

\title{FairTest: Search-Based Fairness Testing for Multi-Agent Reinforcement Learning Systems}

\author{Xiaotong Wang}
\email{1230028459@student.must.edu.mo}
\orcid{0009-0008-4064-1167}
\affiliation{
  \institution{School of Computer Science and Engineering, Macau University of Science and Technology}
  \city{Macao SAR}
  \country{China}
  \postcode{999078}
}

\author{Xuan Xie}
\authornote{Corresponding author.}
\email{xiexuan@must.edu.mo}
\orcid{0000-0003-3981-8515}
\affiliation{
  \institution{School of Computer Science and Engineering, Macau University of Science and Technology}
  \city{Macao SAR}
  \country{China}
  \postcode{999078}
}

\newenvironment{compactitem}
  {\begin{itemize}
   \setlength{\itemsep}{0pt}
   \setlength{\parsep}{0pt}
   \setlength{\topsep}{2pt}
   \setlength{\partopsep}{0pt}}
  {\end{itemize}}

\definecolor{boxgray}{gray}{0.92}
\newsavebox{\findingboxcontent}

\usepackage{pdflscape}
\begin{document}

\begin{abstract}
Multi-agent Reinforcement Learning (MARL) trains a team of agents that share one environment and learn their policies together.
Training maximizes the team return, and a high return does not imply that the rewards are shared fairly among the agents in every episode.
Testing is an established way to discover the failures of deep reinforcement learning, yet few methods address the fairness of MARL.
In this work, we propose FairTest, a search-based testing approach that seeks the unfair executions of a MARL policy.
The design combines search guidance with test prioritization.
The guidance scores each candidate with three fitness functions. One measures the fairness of the runs already performed, another predicts the fairness from abstract states and fairness features, and the third reads the decision uncertainty from the policy.
Crossover and mutation derive further candidates from the observed executions.
The prioritization ranks the candidates by the predicted fairness and the decision uncertainty, so that the runs reach the candidates where failures are expected.
FairTest is evaluated on three environments and two MARL algorithms, and four baselines are given the same budget.
It detects the most fairness failures compared to three baselines with statistical significance and large effect sizes.
The failure count exceeds that of the strongest baseline by 221\% on average and coverage improves by an average of 23\%.
\end{abstract}



\maketitle

\section{Introduction}
\label{sec:introduction}

Multi-agent Reinforcement Learning (MARL) is a research field that studies learning in systems composed of multiple interacting decision makers, and it has become a standard setting for tasks that require several agents to act together~\cite{tan1993multi,lowe2017multi}.
The field combines Reinforcement Learning with game theory and multi-agent systems~\cite{albrecht2024marl}, and it has developed solution concepts of its own for the interaction of learning agents.
Such teams are deployed for traffic signal control~\cite{wei2019colight}, mixed traffic control at unsignalized intersections~\cite{wang2024mixedtraffic}, multi-robot task allocation in warehouses~\cite{pal2025mrtagent}, energy management~\cite{fonseca2024energaize}, video streaming~\cite{subhan2026edge360}, and financial trading~\cite{bao2019fairness}.

The goal of MARL training is to maximize the team return, which is the sum of the rewards that the individual agents receive over an episode.
However, a high return does not imply that these rewards are evenly distributed among the members.
For example, a fleet of service robots can complete every request while one robot remains idle for the whole episode.
Such an allocation is unfair, since it wastes the capacity of the neglected member, concentrates the workload on the remaining agents, and lowers the service quality for the stakeholders.
Therefore, \emph{Fairness} is a practical requirement for MARL policies~\cite{nist2023airmf,reuel2024fairness,ekpo2025fairskillmarl,lamalfa2025fairppo}.

Existing efforts for improving fairness focus on \emph{model training}.
They either add a fairness constraint to the learning objective~\cite{jabbari2017fairness,siddique2020learning,ekpo2025adafair,alqithami2026ecofair,wei2026integrated}, or learn the policy together with an even reward distribution among the agents~\cite{zhang2014fairness,jiang2019learning,zimmer2021learning,grupen2022fairness}.
Nevertheless, both alternatives act on the states visited during training, so unfair executions in unseen states cannot be prevented or detected.
There is also work on performing \emph{testing}, \ie, finding defects, for the deep reinforcement learning agent~\cite{sunba2025testing,zolfagharian2023search,pang2022mdpfuzz,wang2023fuzzing,ilahi2021challenges}.
However, they are designed for the case of the single agent.
In multi-agent systems, the only testing method is MASTest, whose oracle targets functional failures of the team~\cite{ma2024enhancing}, rather than fairness.
To the best of our knowledge, few existing methods are designed for this property, so detecting unfair executions remains a largely untouched problem.

Therefore, this paper proposes \method, a search-based testing approach for fairness in MARL.
The approach searches over the candidate executions of the trained policy for the ones that could be potentially unfair.
The design of \method predominantly rests on two components, \emph{search guidance} and \emph{test prioritization}.
To decide which candidates are worth a run, the guidance evaluates them in advance with three fitness functions, the measured fairness, the predicted fairness, and the decision uncertainty of the policy.
The measured fairness is the only one that requires a run, while the predicted fairness is estimated by a predictor learned from abstract states and from fairness features, and the decision uncertainty is derived from the policy.
Crossover and mutation produce the new candidates as variants of the observed executions, which carries the search to executions that have not been observed yet.
Moreover, a Pareto-based prioritization~\cite{deb2002fast} ranks the candidates by the predicted fairness and the decision uncertainty of the policy, so that the executions are directed to the candidates where failures are expected.
Once a candidate has been run, it is recorded as a failure when the fairness observed in its execution falls below the threshold.
Across generations, the genetic search repeats this cycle and accumulates the unfair executions that it confirms.

An extensive experimental evaluation is conducted. 
Three testing environments (Multi-Agent Particle Environment (MPE)~\cite{lowe2017multi}, Level-Based Foraging (LBF)~\cite{papoudakis2020benchmarking} and Predator-Prey~\cite{lowe2017multi}), two MARL algorithms (QMIX~\cite{rashid2020monotonic} and IQL~\cite{tan1993multi}) and four baselines (Random, GMT~\cite{li2023generative}, MASTest~\cite{ma2024enhancing} and STARLA~\cite{zolfagharian2023search}) are used for comparison.
Across all environment-algorithm configurations, \method detects more fairness failures than every baseline, with statistical significance.
On MPE, LBF, and Predator-Prey, the improvement over the strongest baseline is 106\%, 213\%, and 412\% for IQL and 43\%, 205\%, and 345\% for QMIX.
Coverage also improves over the existing methods on average, with its largest margin on Predator-Prey, where the increase reaches 56\% and 81\%.

The main contributions of this work are summarized as follows.
\begin{compactitem}
    \item \method is proposed as a search-based testing framework for MARL. The framework runs a genetic search for unfair executions efficiently within a limited number of runs and then chooses the next runs economically with a Pareto-based prioritization over the predicted fairness and the decision uncertainty of the policy. 
    \item \method is evaluated on three environments and two MARL algorithms against four baselines under the same testing budget. The experimental results indicate that our approach detects more fairness failures than every baseline in all environment-algorithm configurations.
    \item The source code of \method is released to facilitate replication.
\end{compactitem}

The remainder of this paper is organized as follows. 
The preliminaries are introduced in Section~\ref{sec:preliminaries}, and Section~\ref{sec:method} describes the proposed approach. The experimental evaluation is reported in Section~\ref{sec:experiment}, and Sections~\ref{sec:threats} and~\ref{sec:related_work} discuss the threats to validity and the related work, respectively. 
Section~\ref{sec:conclusion} concludes the paper.

\section{Preliminaries}
\label{sec:preliminaries}

This section introduces the formal model and the fairness index that the proposed approach builds on.
The Reinforcement Learning setting is defined in Section~\ref{subsec:rl}, and Section~\ref{subsec:marl} extends it to MARL.
Section~\ref{subsec:fairness} defines the fairness index, the fairness failure, and the distinction between bias and unfairness.

\subsection{Reinforcement Learning}
\label{subsec:rl}

Reinforcement Learning (RL)~\cite{sutton1998reinforcement} trains an agent to make a sequence of decisions through trial and error, and the agent maximizes the reward that it accumulates from the environment.

\begin{definition}[Markov Decision Process]
A Reinforcement Learning problem is formalized as a Markov Decision Process (MDP)~\cite{bellman1966dynamic}, a tuple
\begin{equation}
M = \langle S, A, T, R, \gamma \rangle,
\label{eq:mdp}
\end{equation}
where $S$ and $A$ are the state space and the action space of the agent, $T: S \times A \times S \to [0, 1]$ is the transition function such that $T(s', a, s)$ gives the probability of reaching state $s'$ by taking action $a$ in state $s$, $R: S \times A \to \mathbb{R}$ assigns the immediate reward of taking an action in a state, and $\gamma \in [0, 1]$ weights the future rewards against the immediate reward.
A policy $\pi: S \to A$ selects an action for every state, and it is the solution that the RL algorithm learns.
The agent starts from an initial state $s_0$ and, at every step $t$, it takes the action $a_t = \pi(s_t)$, receives the reward $r_t = R(s_t, a_t)$, and moves to the next state $s_{t+1}$ sampled from $T$.
\end{definition}

\begin{definition}[Episode]
A single-agent episode is written as $[(s_t, a_t, r_t)]_{t=0}^{T-1}$, with $s_t$ the state at step $t$, $a_t$ the action, and $r_t$ the reward.
The reward accumulated over the episode is $\sum_{t=0}^{T-1} r_t$.
An episode ends when the agent reaches a terminal state or when the time limit is reached.
\end{definition}

\subsection{Multi-Agent Reinforcement Learning}
\label{subsec:marl}

Multi-agent Reinforcement Learning (MARL) extends the single-agent setting (Section~\ref{subsec:rl}) to a team of agents that share one environment and learn their policies at the same time~\cite{lowe2017multi}.
The next state of the environment and the reward of every agent are typically produced by the actions of the whole team.
The environment of an agent is also shaped by the changing policies of the other agents, and the learning problems of the team stay coupled throughout the learning process.
The formal model below represents it with the joint state, the joint action, and the individual rewards of the agents.

\begin{definition}[MARL]
A MARL problem with $N$ agents is written as a tuple
\begin{equation}
\mathcal{M} = \langle \mathcal{S}, \mathcal{A}, \mathcal{T}, \mathcal{R}, \gamma \rangle,
\label{eq:marl}
\end{equation}
where $\mathcal{S}$ is the joint state space of the team, $\mathcal{A} = \mathcal{A}_1 \times \cdots \times \mathcal{A}_N$ combines the action spaces of the individual agents, $\mathcal{T}: \mathcal{S} \times \mathcal{A} \times \mathcal{S} \to [0, 1]$ determines the probability of the next joint state from a joint state and a joint action, $\mathcal{R}$ returns an individual reward $R_i(\mathbf{s}, \mathbf{a})$ for every agent $i$, and $\gamma \in [0, 1]$ is the discount factor.
\end{definition}

The interaction of the team is executed step by step under the model above.
At step $t$, the team is in the joint state $\mathbf{s}_t$, the agents take the joint action $\mathbf{a}_t = (a_t^1, \dots, a_t^N)$, and the individual rewards received at this step are collected in $\mathbf{r}_t = (r_t^1, \dots, r_t^N)$.
The decisions of the team are produced by a joint policy, which is defined below.

\begin{definition}[Joint policy]
A joint policy $\pi: \mathcal{S} \to \mathcal{A}$ selects a joint action for every joint state, and it is the solution that a MARL algorithm learns.
The team starts from an initial joint state $\mathbf{s}_0$ and, at every step $t$, it takes the joint action $\mathbf{a}_t = \pi(\mathbf{s}_t)$, every agent receives its individual reward $r_t^i$, and the environment moves to the next joint state $\mathbf{s}_{t+1}$.
\end{definition}

Executing the joint policy from an initial joint state yields a joint episode, which is the object analyzed by the proposed approach.

\begin{definition}[Joint episode]
\label{def:episode}
A joint episode is a sequence of transitions
\begin{equation}
e = \left[ \left( \mathbf{s}_t, \mathbf{a}_t, \mathbf{r}_t \right) \right]_{t=0}^{T-1},
\label{eq:joint-episode}
\end{equation}
with $T$ the length of the episode.
An episode is generated by executing the joint policy from an initial state until the episode terminates.
Agent $i$ accumulates $x_i(e) = \sum_{t=0}^{T-1} r_t^i$ over the episode.
\end{definition}

\noindent\textbf{A running example.}\quad
The Predator-Prey environment~\cite{lowe2017multi} instantiates the MARL problem defined above.
The team of the environment is composed of three predators that pursue a single prey, and the joint
state of the environment records the positions of the four agents.
The action space of a predator contains the movement directions, and the joint action of the team
combines the directions that the three predators choose at a step.
The transition of the environment moves the predators along the chosen directions, and the reward of
a predator equals the negative distance between its own position and the position of the prey.
A joint policy of the team selects a joint action for the current joint state, and executing this
policy over one chase produces a joint episode, in which every predator accumulates the rewards of
its own steps.
The three predators of one chase usually accumulate different rewards, and the fairness of this
difference is the subject of Section~\ref{subsec:fairness}.

\subsection{Fairness in MARL}
\label{subsec:fairness}

Fairness concerns how a resource or an outcome is divided among a set of individuals~\cite{mehrabi2021survey},
since a division determines what each individual receives.
In MARL, the quantity that an agent receives is its individual reward, and the fairness of the team is
typically evaluated on how evenly the rewards of the agents are distributed~\cite{jiang2019learning,zimmer2021learning}.
There are numerous fairness definitions, such as utilitarian criteria~\cite{moulin2003fairdivision},
Rawlsian criteria~\cite{rawls1971theory,zhang2014fairness}, envy-free division~\cite{varian1974equity},
and multi-objective formulations~\cite{jiang2019learning,zimmer2021learning}.
In this work, we adopt Jain's Fairness Index (JFI)~\cite{jain1984quantitative} as the fairness measurement.
JFI measures the evenness of a reward distribution and is a classical measure that has been widely used to quantify the fairness of resource allocation in multi-agent systems~\cite{subhan2026edge360}.

\noindent\textbf{Jain's Fairness Index (JFI).}\quad
JFI is defined as
\begin{equation}
\mathrm{JFI}(e) = \frac{\left( \sum_{i=1}^{N} x_i(e) \right)^{2}}{N \cdot \sum_{i=1}^{N} x_i(e)^{2}},
\label{eq:jfi}
\end{equation}
where $e$ is an episode, and $x_i(e)$ is the total reward of agent $i$ over the episode.
The index lies in the interval $[1/N, 1]$.
The value $1$ means that all agents accumulate the same reward, and the value $1/N$ means that a single agent receives the whole reward.
The intermediate values describe the partial degrees of inequality.

\noindent\textbf{Unfairness and Bias.}\quad
In machine learning, bias denotes a systematic deviation in the data or in the algorithm that skews the decisions toward certain groups, and the unequal outcomes caused by this deviation are described as unfairness~\cite{mehrabi2021survey}.
The MARL fairness literature~\cite{bao2019fairness,subhan2026edge360,grupen2022fairness,zimmer2021learning} uses unfairness in the same outcome sense for the unequal allocation of rewards among agents, and this convention is adopted in this work.

\begin{definition}[Fairness failure]
\label{def:fairness-failure}
A fairness failure occurs during the execution of an episode $e$ when $\mathrm{JFI}(e) \le \theta$ for a failure threshold $\theta$.
\end{definition}

\section{Methodology}
\label{sec:method}

This section presents \method, a search-based approach for the fairness testing of MARL policies.
The approach is a genetic algorithm that evolves a population of joint episodes toward unfair ones, and a fairness oracle identifies the unfair episodes by the criterion of Definition~\ref{def:fairness-failure}.
The prioritization spends the testing budget on the most promising candidates.
The remainder of this section details each component.

\subsection{Reformulation as a Search Problem}

A joint episode of a trained policy is a finite sequence of transitions, as defined in Definition~\ref{def:episode}.
Each execution of an episode is traced step by step, and the trace of every step keeps the joint state, the joint action, the individual rewards, and the per-agent Q-values.
The testing problem is formulated as a multi-objective search problem over the space of joint episodes, with the following elements.

\begin{compactitem}
    \item \textbf{Individual.} An individual is a joint episode. An episode is preferred when the trained policy distributes the accumulated rewards among the agents in an unfair manner.
    \item \textbf{Initial population.} The initial population is a set of joint episodes generated by random executions of the trained policy and filtered by the Pareto-based prioritization described in Section~\ref{subsec:prioritization}.
    \item \textbf{Operators.} Crossover and mutation generate offspring from the current population, and a MOSA-based selection~\cite{panichella2015reformulating} determines the survivors of each generation (Section~\ref{subsec:operators}).
    \item \textbf{Fitness functions.} Three fitness functions are minimized during the search: the fairness index, the predicted probability of fairness, and the decision uncertainty (Section~\ref{subsec:fitness}).
    \item \textbf{Termination criteria.} The search is bounded by a fixed testing budget. One run consists of three rounds of three generations, and the budget is fixed to 18{,}000 episodes per run.
\end{compactitem}

\subsection{Overview of the Approach}
\label{subsec:overview}

The approach follows the classic loop of a genetic algorithm: an initial population is built, offspring are generated from it, and a selection step decides which episodes survive to the next generation.
Every episode is evaluated on the three fitness functions (Section~\ref{subsec:fitness}).
The initial population and the offspring are filtered by a Pareto-based prioritization, which keeps the episodes that are most promising for fairness testing.
Throughout the search, the episodes that the fairness oracle judges as unfair are collected in a failure pool.
The search stops when the fixed testing budget is used up.
\begin{figure}[htbp]
  \centering
  \includegraphics[width=0.95\linewidth]{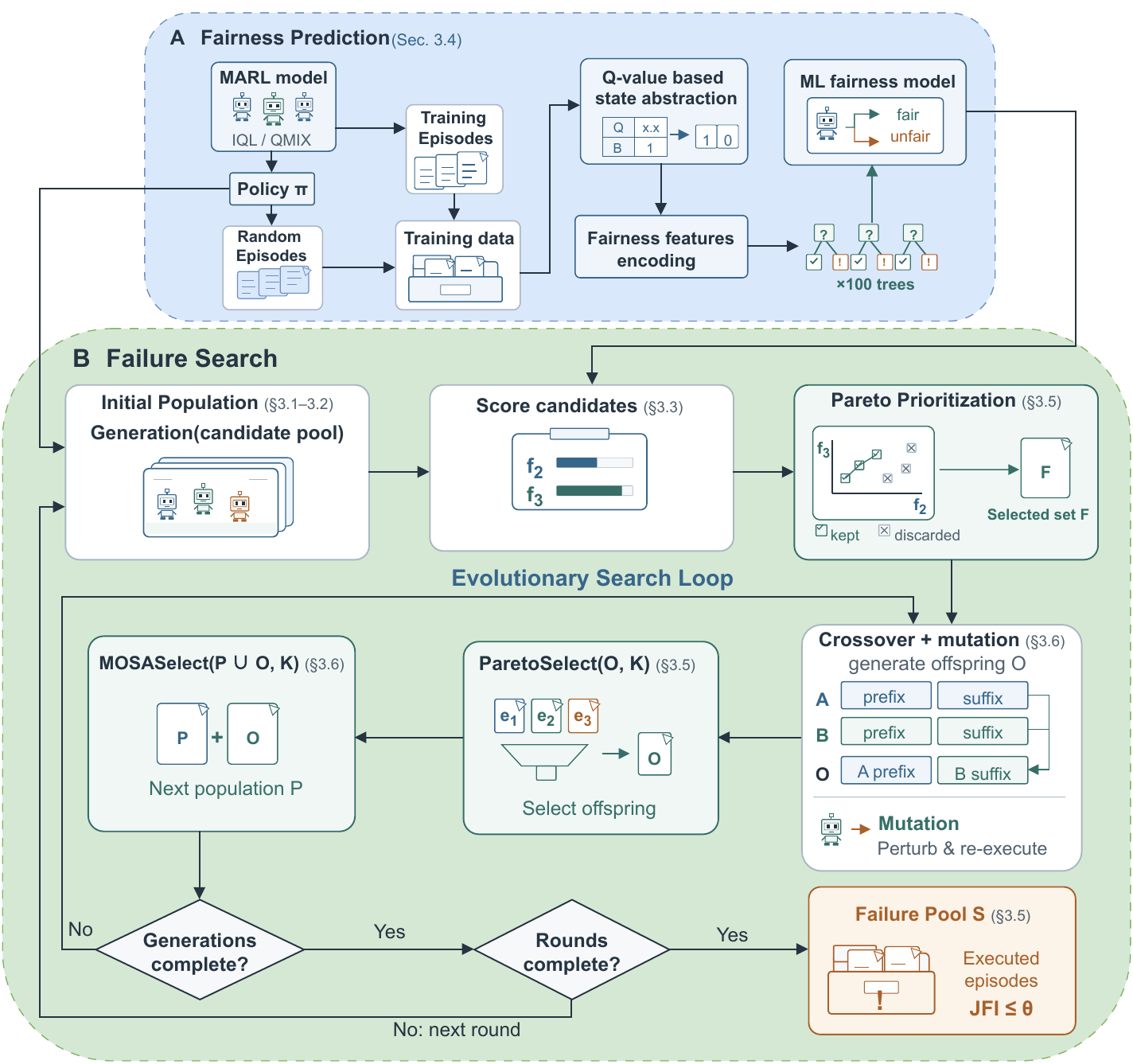}
  \vspace{-10pt}
  \caption{Overview of \method.}
  \label{fig:overview}
\end{figure}
Algorithm~\ref{alg:pgfair} implements this process, and the main steps of one run are listed below.
\begin{enumerate}
    \item Random executions of the trained policy generate a candidate pool, and episodes shorter than five steps are discarded (Section~\ref{subsec:prioritization}, line 3).
    \item The predicted fairness $f_2$ and the decision uncertainty $f_3$ are evaluated on every candidate, and the Pareto-based prioritization keeps the most promising candidates as the initial population (Sections~\ref{subsec:fitness} and \ref{subsec:prioritization}, lines 4-5).
    \item The initial population is re-executed from its saved seeds with the environment states recorded at every step, and the three fitness functions are evaluated on it, which makes it the current population of the first generation (Section~\ref{subsec:fitness}, lines 6-7).
    \item At every generation, crossover and mutation produce offspring from the current population, the mutated offspring are executed in the environment, and the fitness values are evaluated on the offspring (Sections~\ref{subsec:operators} and \ref{subsec:fitness}, lines 8-10).
    \item The prioritization is applied to the offspring, and the selection step merges the surviving offspring with the parent population to form the population of the next generation (Sections~\ref{subsec:prioritization} and \ref{subsec:operators}, lines 11-12).
    \item Every executed episode is checked against the failure threshold, and the episodes whose fairness index falls below the threshold are added to the failure pool (Sections~\ref{subsec:fitness} and \ref{subsec:prioritization}, lines 7 and 10).
\end{enumerate}
The following subsections describe the fitness functions, the fairness predictor, the prioritization, and the search operators in detail.

\begin{algorithm}[!htbp]
\caption{\method}
\label{alg:pgfair}
\begin{algorithmic}[1]
\Require trained policy $\pi$, fairness predictor $\mathcal{M}$, rounds $R$, generations $G$, failure threshold $\theta$
\Ensure failure pool $\mathcal{S}$ of fairness failure episodes
\State $\mathcal{S} \gets \emptyset$
\For{$r \gets 1$ \textbf{to} $R$}
    \State $\mathcal{C} \gets$ generate candidates by random executions of $\pi$
    \State score $\mathcal{C}$ on $f_2$ and $f_3$
    \State $P \gets \textsc{ParetoSelect}(\mathcal{C}, K)$
    \State re-execute $P$ and record the environment states
    \State evaluate $f_1, f_2, f_3$ on $P$ and update $\mathcal{S}$
    \For{$g \gets 1$ \textbf{to} $G$}
        \State $O \gets$ generate offspring from $P$ by crossover and mutation
        \State execute the mutated offspring in the environment, evaluate $f_1, f_2, f_3$ on $O$, and update $\mathcal{S}$
        \State $O \gets \textsc{ParetoSelect}(O, K)$
        \State $P \gets \textsc{MOSASelect}(P, O, K)$
    \EndFor
\EndFor
\State \Return $\mathcal{S}$
\end{algorithmic}
\end{algorithm}

\subsection{Fitness Functions}
\label{subsec:fitness}

Three fitness functions are used to guide the search toward unfair episodes.
The fairness index directly captures the inequality of the outcome.
The predicted fairness estimates the probability of a fair outcome from evidence beyond the accumulated rewards.
The decision uncertainty identifies episodes in which a small perturbation is likely to change the outcome.

\subsubsection{Fairness index ($f_1$).}

The first fitness function is Jain's Fairness Index~\cite{jain1984quantitative} of the accumulated individual rewards, as defined in Section~\ref{sec:preliminaries}.
The fitness of an episode is
\begin{equation}
f_1(e) = \mathrm{JFI}(e),
\label{eq:f1}
\end{equation}
where $\mathrm{JFI}(e)$ is given by Equation~\ref{eq:jfi} and the accumulated rewards $x_i(e)$ are defined in Definition~\ref{def:episode}.
During the search, $f_1$ is computed from the actual rewards of the executed episodes, and it is available whenever a candidate or a mutated offspring is run in the environment.
An episode $e$ is a fairness failure if and only if $f_1(e)$ does not exceed the failure threshold $\theta$.

\textbf{Example 1:} 
The Predator-Prey (PP) environment is used as an example, in which three predators chase a single prey on a two-dimensional continuous plane.
The prey is caught when any predator comes into contact with it, which occurs when the Euclidean distance between the centers of the two falls below the sum of their radii. 
At every step, a predator chooses a movement direction from its perception of the prey, and the reward of that step is the negative Euclidean distance to the prey.
Each episode is one complete chase, which starts from an initial state and ends when the time limit is reached.
When two of the three predators chase the prey together and the third one lags far behind, the two participants keep their rewards near zero, whereas the third one accumulates a large negative amount.
The fairness index of the episode then falls far below the failure threshold, and the episode is a fairness failure.
Minimizing the fairness index drives the search toward such episodes, which are promising candidates for the search operators, because their variants from crossover and mutation are more likely to contain similar failures.

\subsubsection{Predicted fairness ($f_2$).}

The second fitness function is the probability of fairness predicted by the machine-learned predictor of Section~\ref{subsec:ml}.
The fairness index of an episode is known only after the episode is executed to the end, and the predictor is trained before the search to estimate this value from the recorded transitions without a new execution.
The predictor thereby scores a candidate episode at a much lower cost, and it can anticipate the unfairness of an episode whose outcome has not yet been observed.
Let $P(\mathit{fair} \mid e)$ be the probability assigned to the fair class by the predictor for episode $e$.
The fitness is defined as
\begin{equation}
f_2(e) = P(\mathit{fair} \mid e).
\label{eq:f2}
\end{equation}
Minimizing $f_2$ is equivalent to maximizing the predicted probability of unfairness, which directs the search toward the episodes that the predictor considers most likely to be unfair.

\textbf{Example 2:} In the Predator-Prey environment, a predator may fall behind the pursuit and keep a larger distance to the prey, so it accumulates a lower reward.
The accumulated rewards can still stay close when the trailing predator rejoins the chase near the end, and the fairness index of such an episode then looks normal, although the predator takes little part in the capture.
The single number of the fairness index cannot reveal this pattern, whereas the fairness features preserve the outcome in detail.
The predictor is trained on labeled episodes, and it learns the combinations of abstract states and fairness features that are typically paired with unfair outcomes.
When the features of a candidate episode match a memorized combination, the predictor flags the episode as unfair, even if its accumulated rewards look balanced (Section~\ref{subsec:ml}).

\subsubsection{Decision uncertainty ($f_3$).}

The third fitness function measures the uncertainty of the policy decisions within an episode with the DeepGini measure~\cite{feng2020deepgini}, which is proposed to locate the decisions of a deep learning model that are most likely to go wrong from its output probabilities.
A fairness failure is often triggered at a decision boundary, where the policy is nearly indifferent between several actions.
A small perturbation of the environment at such a step can change the selected action and the resulting reward allocation, and the changed outcome may expose a fairness failure that the original episode does not contain.
These episodes are the most promising targets for the search operators.
Let $p_{ik}^t$ be the softmax probability of action $k$ for agent $i$ at step $t$, obtained from the $Q$-values.
The fitness is
\begin{equation}
f_3(e) = \frac{1}{|\mathcal{T}(e)|} \cdot \sum_{(t,i) \in \mathcal{T}(e)} \sum_{k=1}^{A} \left( p_{ik}^t \right)^2,
\label{eq:f3}
\end{equation}
where $\mathcal{T}(e)$ is the set of agent-step pairs in $e$ and $A$ is the number of actions.
During the search, $f_3$ is computed from the $Q$-values recorded in the executed episodes, and it requires no additional execution.
The average over $\mathcal{T}(e)$ makes the fitness comparable across episodes of different lengths.
The DeepGini measure aggregates the softmax probabilities of all actions, which reflects the overall uncertainty of a decision accurately.
Minimizing $f_3$ is equivalent to maximizing the classic DeepGini measure $1 - \sum_{k=1}^{A} \left(p_{ik}^t\right)^2$, which grows as the probability mass spreads over more actions.
The sign is inverted to keep all three fitness functions in the same minimization direction.

\textbf{Example 3:} Take the Predator-Prey environment as an example.
A predator may hesitate between chasing the prey directly and moving around it to cut off its escape, and the two options can have similar values under the policy, which makes the choice of the predator ambiguous.
This ambiguity leaves no trace in the completed rewards, so the fairness index of the episode may stay high.
The abstract states give no better signal, because the abstraction rounds the option values into coarse bins that discard the closeness of the two values.
The decision uncertainty, in contrast, is computed from the original $Q$-values, where the gap between them is preserved, without relying on the completed rewards or a trained model.
The signal is therefore visible before any unfair allocation has formed, and it marks the hazard earlier than the fairness index or the predictor.
An uncertain step is also the place where a disturbance has the strongest effect, since a small perturbation can flip the selected action, exposing a fairness failure that the original episode does not contain.

\subsection{Fairness Predictor}
\label{subsec:ml}

Executing an episode requires running the trained policy in the environment, and the testing
budget allows only a limited number of such executions.
An estimate of the fairness of an episode that does not require a new execution lets the search
screen the candidates before spending the budget, and the episodes that the estimate marks as
unfair can then be executed for confirmation.
Thus, we leverage a machine learning model as the fairness predictor, which provides such an estimate.
It is trained before the search, and its output serves as the fitness function $f_2$ of Section~\ref{subsec:fitness} and as one of the two objectives of the prioritization in Section~\ref{subsec:prioritization}.
The construction of the predictor proceeds in four steps.
A set of labeled episodes is prepared, and every episode is marked as fair or unfair according to its JFI.
The concrete states of these episodes are mapped to abstract states, which compresses the state space into a compact representation.
Every labeled episode is encoded into a feature vector that combines the abstract-state representation with reward-derived statistics.
A random forest classifier is trained on the encoded episodes, and the probability of the fair class is output as $f_2$.
The following paragraphs detail each step.

\noindent\textbf{Training data.}\quad
The training data of the predictor consist of two parts.
The first part is sampled from the episodes explored during the training of the agents.
An episode that appears at position $i$ in the training sequence is selected with a probability proportional to $i$, so that the episodes from the later training stages, which are closer to the final policy, receive higher weights.
The second part is collected from random executions of the trained policy.
Every episode is labeled as fair if its JFI is not below the failure threshold and as unfair otherwise.

\noindent\textbf{State abstraction.}\quad
The episodes of the training data are sequences of concrete states, and each concrete state records the positions and the velocities of all agents at one step.
Such a state is too high-dimensional to be used as a feature directly, and an episode typically contains a prohibitively large number of states, which are difficult to encode.
The concrete states are therefore compressed into abstract states, following the $Q^*$-irrelevance abstraction~\cite{li2006state}.
This abstraction groups the states according to their optimal $Q$-values.
Each agent is abstracted independently, and the abstraction of agent $i$ is a coarse representation of its $Q$-value vector,
\begin{equation}
\phi_d(\mathbf{q}_i) = \left( \left\lceil \frac{q_i^1}{d} \right\rceil, \dots, \left\lceil \frac{q_i^A}{d} \right\rceil \right),
\label{eq:abs}
\end{equation}
where $\mathbf{q}_i$ is the $Q$-value vector of agent $i$, $d$ is the abstraction level, and the $Q$-values are mapped into buckets of width $d$.
The joint abstract state of a step is the tuple of the individual abstractions of all agents.
Each episode is then encoded as a binary vector over the joint abstract states observed in the training data, where a dimension is set to $1$ if and only if the corresponding joint abstract state appears in the episode.
The abstract states are derived from the $Q$-values and carry no reward information.

\noindent\textbf{Fairness features.}\quad
A machine learning predictor built on the Q-value abstraction alone describes an episode by the set of abstract states that appear in it, which reflects the behavior of the agents but omits the rewards that they accumulate.
These rewards are the key information that this representation lacks, since the JFI of Equation~\ref{eq:jfi} is computed from them.
The fairness features fill this gap by adding the accumulated rewards of the agents and the statistics of their distribution to the input of the predictor.
With this addition, the predictor is aligned with the fairness criterion and can better identify the unfair episodes.
The fairness features consist of two groups.
\begin{compactitem}
\item \textbf{Individual features.} For each agent $i$, two quantities are encoded: the accumulated reward $x_i$ and the reward share $x_i / \sum_{j} x_j$, which is the fraction of the team reward obtained by agent $i$.
\item \textbf{Team-level features.} The team reward distribution is summarized with three statistics of the accumulated rewards.
\begin{compactitem}
        \item \textbf{Coefficient of variation.} $\mathrm{cv}(x_1, \dots, x_N)$ is the ratio of the standard deviation to the mean of the accumulated rewards, and it measures the dispersion relative to the average allocation.
        \item \textbf{Gini coefficient.} $\mathrm{gini}(x_1, \dots, x_N)$ measures the inequality of the allocation on a scale from $0$, where every agent receives an equal reward, to $1$, where a single agent receives the whole reward.
        \item \textbf{Pairwise reward difference.} $|x_i - x_j|$ is computed for every pair of agents and captures the gap between the allocations of two agents.
    \end{compactitem}
\end{compactitem}
Each value is normalized to $[0, 1]$ with a feature-specific scaling and discretized into one-hot buckets.
Algorithm~\ref{alg:features} describes the complete construction, where $\mathrm{bucket}(v)$ maps a normalized scalar $v$ to a $B$-dimensional one-hot vector over $B$ equal-width buckets.

\begin{algorithm}[!htbp]
\caption{Feature construction for the fairness predictor}
\label{alg:features}
\begin{algorithmic}[1]
\Require episode $e$, the set of joint abstract states $\Phi$ observed in the training data, abstraction level $d$, number of buckets $B$
\Ensure feature vector $\mathbf{v}(e)$
\State $\mathbf{v} \gets \mathbf{0}^{|\Phi|}$ \Comment{presence-absence vector over $\Phi$}
\For{each transition $(\mathbf{s}_t, \mathbf{a}_t, \mathbf{r}_t) \in e$}
    \State $\phi \gets$ joint abstract state of $\mathbf{s}_t$ by Equation~\ref{eq:abs}
    \If{$\phi \in \Phi$}
        \State $v[\phi] \gets 1$
    \EndIf
\EndFor
\State compute the accumulated rewards $x_i \gets \sum_t r_t^i$ for every agent $i$
\For{each agent $i$}
    \State $\mathbf{v} \gets \mathbf{v} \oplus \mathrm{bucket}(x_i)$ \Comment{accumulated reward, G1}
    \State $\mathbf{v} \gets \mathbf{v} \oplus \mathrm{bucket}\!\left(\frac{x_i}{\sum_j x_j}\right)$ \Comment{reward share, G1}
\EndFor
\State $\mathbf{v} \gets \mathbf{v} \oplus \mathrm{bucket}\!\left(\mathrm{cv}(x_1, \dots, x_N)\right)$ \Comment{coefficient of variation, G2}
\State $\mathbf{v} \gets \mathbf{v} \oplus \mathrm{bucket}\!\left(\mathrm{gini}(x_1, \dots, x_N)\right)$ \Comment{Gini coefficient, G2}
\For{each pair $(i, j)$ with $i < j$}
    \State $\mathbf{v} \gets \mathbf{v} \oplus \mathrm{bucket}(|x_i - x_j|)$ \Comment{pairwise difference, G2}
\EndFor
\State \Return $\mathbf{v}$
\end{algorithmic}
\end{algorithm}

\noindent\textbf{Classifier.}\quad
The classifier learns the boundary between the fair and the unfair episodes of the labeled data, and it predicts the class of an episode that the training has not seen.
The class of an episode is rarely decided by a single feature, since the abstract-state indicators and the fairness features of the encoding act together.
The encoding is high-dimensional and sparse, and a random forest fits this kind of input well, as a tree reaches its class through a series of checks, with every check examining one feature.
The checks are applied one after another along the path from the root of the tree to a leaf, and the features examined on the path jointly determine the class, which lets a tree express boundaries that combine the two parts of the encoding.
A single tree overfits the particular sample on which it is grown, and the forest avoids this by training many trees and averaging their predictions.
Every tree is trained on a different sample that is drawn from the labeled episodes with repetition, and the averaged prediction stays stable even when the labeled data are limited, which the search needs, as the forest also has to score episodes that are not executed.
When the two classes are heavily imbalanced and the unfair episodes form only a tiny part of the labeled data, the training assigns the unfair episodes a higher weight, so that the forest does not simply predict the fair class for every episode.
The forest outputs the probability of the fair class, and this probability is used as the fitness function $f_2$ of Section~\ref{subsec:fitness} during the search, where a low value marks an episode that is likely to be unfair.

\subsection{Pareto-based Test-Case Prioritization}
\label{subsec:prioritization}

The prioritization decides which candidates are worth the testing budget. 
It operates on the two predictive objectives $f_2$ and $f_3$, and it is applied twice in every round, to the initial candidate pool and to the offspring of every generation. 
$f_2$ anticipates the unfairness of an episode, and $f_3$ reflects the uncertainty of the policy decisions.
Failure determination is independent of the population selection: an episode is a failure if and only if its fairness index falls below the threshold $\theta$, and every evaluated episode is checked against this criterion, whether the prioritization keeps it or not.
A failure is always confirmed by an executed episode rather than by a prediction.
The failure pool $\mathcal{S}$ collects the failures among all evaluated episodes, so a failure is recorded even when its candidate is discarded by the prioritization.
The procedure contains three steps:
\begin{enumerate}
    \item The candidates are pre-filtered on the decision uncertainty.
    \item The Pareto front of $f_2$ and $f_3$ then keeps the survivors.
    \item The crowding distance finally adjusts the survivors to the target size.
\end{enumerate}

The pre-filter considers only the decision uncertainty.
The candidates are sorted by $f_3$ in ascending order, and the most uncertain candidates are kept, up to the size of the candidate pool.
The number of kept candidates is capped at the size of the candidate pool, and the pool is generated at this size under the default setting, so the whole pool survives the pre-filter.
When a much smaller population is retained, as in the budget-ratio study, the cap is smaller than the pool, and the pre-filter restricts the candidates to the most uncertain ones before the Pareto-based filter~\cite{panichella2015reformulating,sayyad2013pareto,epitropakis2015empirical,hu2024test} is applied.

Next, a Pareto-front filter is applied to $f_2$ and $f_3$.
The Pareto front~\cite{pareto1906manual,deb2001multi,coello2007evolutionary,zitzler1999multiobjective} is a standard notion in multi-objective optimization, and it denotes the candidates that represent the best trade-offs between the objectives.
A candidate $c$ dominates another candidate $c'$, denoted by $c \prec c'$, when it is no worse on both objectives and strictly better on at least one.
A dominated candidate brings nothing new to the search: the candidate that dominates it is at least as unfair and at least as uncertain, so the dominated one can be discarded.
For instance, suppose that a candidate has the scores $(0.1, 0.2)$ on $(f_2, f_3)$ and another candidate has $(0.3, 0.4)$.
Both scores are minimized during the search, and the first candidate is lower on both of them, which makes it the better candidate on both dimensions.
The second candidate is dominated and discarded, and it cannot be a member of the Pareto front.
The candidates that are not dominated by any other candidate form the Pareto front,
\begin{equation}
\mathcal{F} = \left\{ c \in \mathcal{C} \;\middle|\; \nexists\, c' \in \mathcal{C} : c' \prec c \right\},
\label{eq:paretofront}
\end{equation}
$\mathcal{F}$ contains the candidates for which no other candidate in $\mathcal{C}$ is better on both objectives.
Every front member is the best in at least one objective, so the front keeps the most-unfair candidates, the most-uncertain candidates, and the trade-offs between them.

Finally, the crowding distance is used to adjust the set to the target size $K$.
The crowding distance of a candidate $c$ is the sum, over the two objectives, of the normalized distances to its two neighbors in the sorted order,
\begin{equation}
d(c) = \sum_{i \in \{2, 3\}} \frac{f_i(c^{+}) - f_i(c^{-})}{f_i^{\max} - f_i^{\min}},
\label{eq:crowding}
\end{equation}
where $c^{+}$ and $c^{-}$ are the neighbors of $c$ in the ascending order of $f_i$, and the boundary candidates receive an infinite distance.
That is to say, when the Pareto front contains fewer than $K$ candidates, the remaining candidates with the largest crowding distances are appended.
When it contains more than $K$, the front is pruned to the $K$ candidates with the largest crowding distances.
Algorithm~\ref{alg:pareto} formalizes the procedure.
The crowding distance spreads the kept candidates over the score space, which lets the population cover different regions of the space instead of crowding around a single point.
The prioritization keeps the candidates that are predicted to be unfair, and it discards the candidates that are predicted to be fair.

\begin{algorithm}[!htbp]
\caption{Pareto-based test-case prioritization (\textsc{ParetoSelect})}
\label{alg:pareto}
\begin{algorithmic}[1]
\Require candidate set $\mathcal{C}$ with scores $(f_2, f_3)$, target size $K$
\Ensure selected subset of size $K$
\State sort the candidates by $f_3$ in ascending order
\State $\mathcal{C}' \gets$ the most uncertain candidates \Comment{most uncertain}
\State $\mathcal{F} \gets$ non-dominated candidates of $\mathcal{C}'$ on $(f_2, f_3)$
\If{$|\mathcal{F}| < K$}
    \State compute the crowding distances of $\mathcal{C}' \setminus \mathcal{F}$ on $(f_2, f_3)$
    \State append candidates with the largest crowding distances until the target size is reached
\ElsIf{$|\mathcal{F}| > K$}
    \State compute the crowding distances of $\mathcal{F}$ on $(f_2, f_3)$
    \State keep the $K$ candidates with the largest crowding distances
\EndIf
\State \Return $\mathcal{F}$
\end{algorithmic}
\end{algorithm}

\subsection{Search Operators}
\label{subsec:operators}

\noindent\textbf{Crossover.}\quad
Crossover combines two joint episodes into two offspring, and the two parents are obtained in two different ways.
The first parent is drawn by tournament selection~\cite{miller1996genetic}: a small group of episodes is drawn at random from the population, and the fittest episode of the group is chosen, which gives weaker episodes a chance to be selected and preserves the diversity of the population.
A crossover point $\ell$ is then chosen with equal probability in this parent.
An episode is the joint trajectory of the whole team, and the joint abstract state of a step is the tuple of the individual abstractions of all agents.
The joint abstract state at the crossover point is computed, and a second episode that contains the same joint abstract state at some step is located in the population, which serves as the other parent, so that the two episodes are joined at a step at which the whole team is in an equivalent situation.
If no such episode is found, the procedure restarts.
The two episodes are split at the crossover point, and the resulting parts are recombined crosswise.
The parent episodes are: 
\begin{equation}
e_A = \left[(\mathbf{s}_1, \mathbf{a}_1, \mathbf{r}_1), \ldots, \boldsymbol{(\mathbf{s}_\ell, \mathbf{a}_\ell, \mathbf{r}_\ell)}, \ldots, (\mathbf{s}_T, \mathbf{a}_T, \mathbf{r}_T)\right],
\label{eq:crossover-source}
\end{equation}
\begin{equation}
e_B = \left[(\mathbf{s}'_1, \mathbf{a}'_1, \mathbf{r}'_1), \ldots, \boldsymbol{(\mathbf{s}'_\ell, \mathbf{a}'_\ell, \mathbf{r}'_\ell)}, \ldots, (\mathbf{s}'_T, \mathbf{a}'_T, \mathbf{r}'_T)\right],
\end{equation}
where $\mathbf{s}_t = (s_t^1, \ldots, s_t^N)$, $\mathbf{a}_t = (a_t^1, \ldots, a_t^N)$, and $\mathbf{r}_t = (r_t^1, \ldots, r_t^N)$ are the joint state, the joint action, and the individual rewards of the $N$ agents at step $t$, and the bold transition at position $\ell$ is the crossover point.
The first offspring inherits the prefix of the parent up to the crossover point and the suffix of the matching episode from the crossover point,
\begin{equation}
o_1 = \left[(\mathbf{s}_1, \mathbf{a}_1, \mathbf{r}_1), \ldots, \boldsymbol{(\mathbf{s}'_\ell, \mathbf{a}'_\ell, \mathbf{r}'_\ell)}, \ldots, (\mathbf{s}'_T, \mathbf{a}'_T, \mathbf{r}'_T)\right],
\end{equation}
and the second offspring inherits the complementary parts,
\begin{equation}
o_2 = \left[(\mathbf{s}'_1, \mathbf{a}'_1, \mathbf{r}'_1), \ldots, \boldsymbol{(\mathbf{s}_\ell, \mathbf{a}_\ell, \mathbf{r}_\ell)}, \ldots, (\mathbf{s}_T, \mathbf{a}_T, \mathbf{r}_T)\right].
\end{equation}
The offspring are truncated to the episode length limit.
The matching on the abstract state increases the chance that an offspring is valid, but a spliced sequence may not be an actual trajectory, since the suffix starts from a state that the first part does not reach.
Such offspring are not executed during the search, and their validity is confirmed once the search terminates.
During the search, they are scored with the predicted fairness and the decision uncertainty, which are computed from the recorded transitions of the parents.

\noindent\textbf{Mutation.}\quad
Mutation perturbs an episode at a randomly chosen step to explore nearby states.
An episode is drawn by tournament selection, and a mutation point $\ell$ is chosen uniformly at random.
The environment is restored to the recorded state at the mutation point.
The positions and velocities of the agents under test are scaled by independent uniform factors and clipped to the environment bounds.
Only the agents under test are perturbed, and the other entities of the environment remain unchanged.
The episode is re-executed from the mutation point, and the recomputed transitions replace the transitions after the mutation point.
Let $e$ be the selected episode and $e'$ the mutated episode,
\begin{equation}
e = \left[(\mathbf{s}_1, \mathbf{a}_1, \mathbf{r}_1), \ldots, \boldsymbol{(\mathbf{s}_\ell, \mathbf{a}_\ell, \mathbf{r}_\ell)}, \ldots, (\mathbf{s}_T, \mathbf{a}_T, \mathbf{r}_T)\right],
\end{equation}
\begin{equation}
e' = \left[(\mathbf{s}_1, \mathbf{a}_1, \mathbf{r}_1), \ldots, \boldsymbol{(\tilde{\mathbf{s}}_\ell, \tilde{\mathbf{a}}_\ell, \tilde{\mathbf{r}}_\ell)}, \ldots, (\tilde{\mathbf{s}}_T, \tilde{\mathbf{a}}_T, \tilde{\mathbf{r}}_T)\right],
\label{eq:mutation}
\end{equation}
where the bold transition is the mutation point, and the tilded transitions are produced by the re-execution of the joint policy.
The mutated episode joins the population.
The re-execution lets the whole team react to the perturbed state, and the recomputed actions of the agents determine a new reward allocation, which can expose fairness failures that the original episode does not contain.

\noindent\textbf{Selection.}\quad
Selection determines the population of the next generation, and it is performed by the Many-Objective Sorting Algorithm (MOSA)~\cite{panichella2015reformulating}.
MOSA is adopted to minimize the three fitness functions at the same time, and its ranking gives priority to the fitness functions that are not yet covered, which prevents the search from staying on the goals that have already been satisfied.
The three fitness functions are computed from the individual rewards of the team, and the ranking is aligned with the team-level search goals.

The selection works as follows.
The parent population and the surviving offspring are first merged into a single set, and the value of every candidate on the three fitness functions is evaluated.
A preference-based ranking is then applied to the merged set: for every fitness function that is not yet covered, the candidate with the best value on that fitness function is placed at the top, and the remaining candidates are sorted into non-dominated fronts.
The fronts are filled into the new population one after another, and the candidates within a front are ordered by their crowding distance, which keeps the selected population covering the best solutions in every direction and well spread in the search space.
Any evaluated episode whose fairness index does not exceed the threshold $\theta$ is added to the failure pool $\mathcal{S}$ of Algorithm~\ref{alg:pgfair}.

\section{Experimental Evaluation}
\label{sec:experiment}

\subsection{Experimental Setup}

\begin{figure}[htbp]
  \centering
  \begin{subfigure}[b]{0.46\linewidth}
    \centering
    \includegraphics[width=\linewidth]{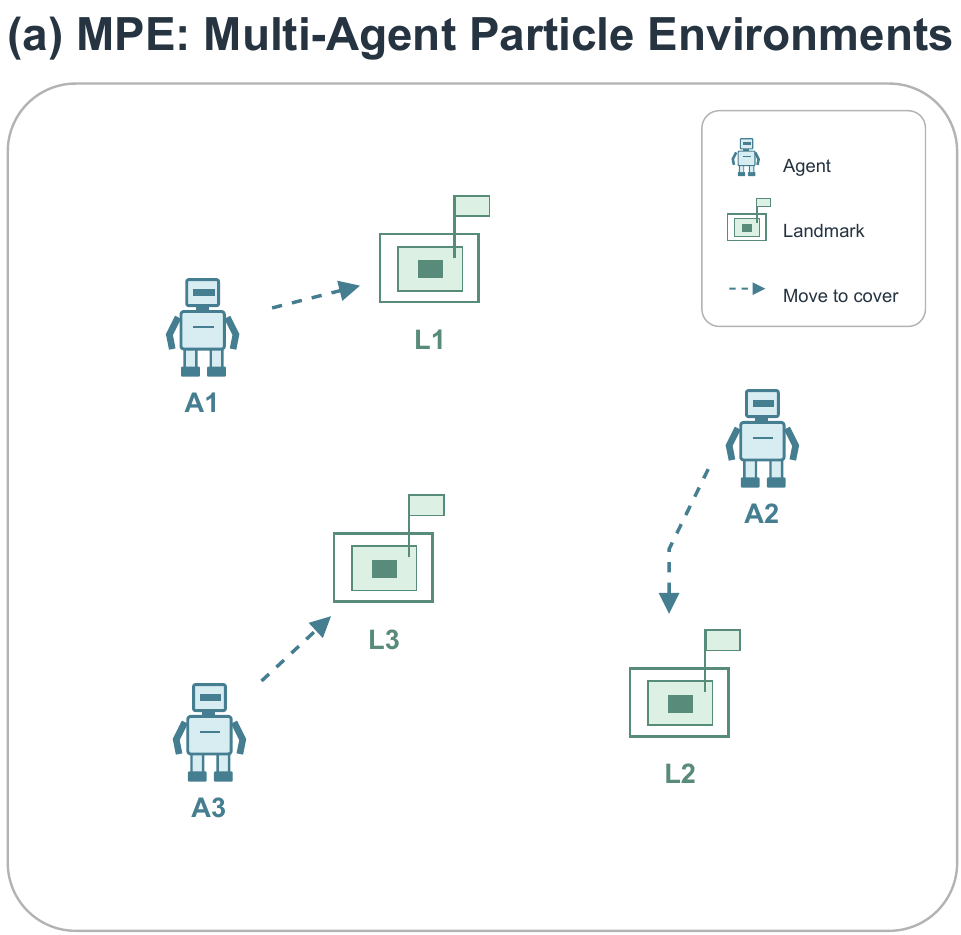}
  \end{subfigure}
  \hfill
  \begin{subfigure}[b]{0.46\linewidth}
    \centering
    \includegraphics[width=\linewidth]{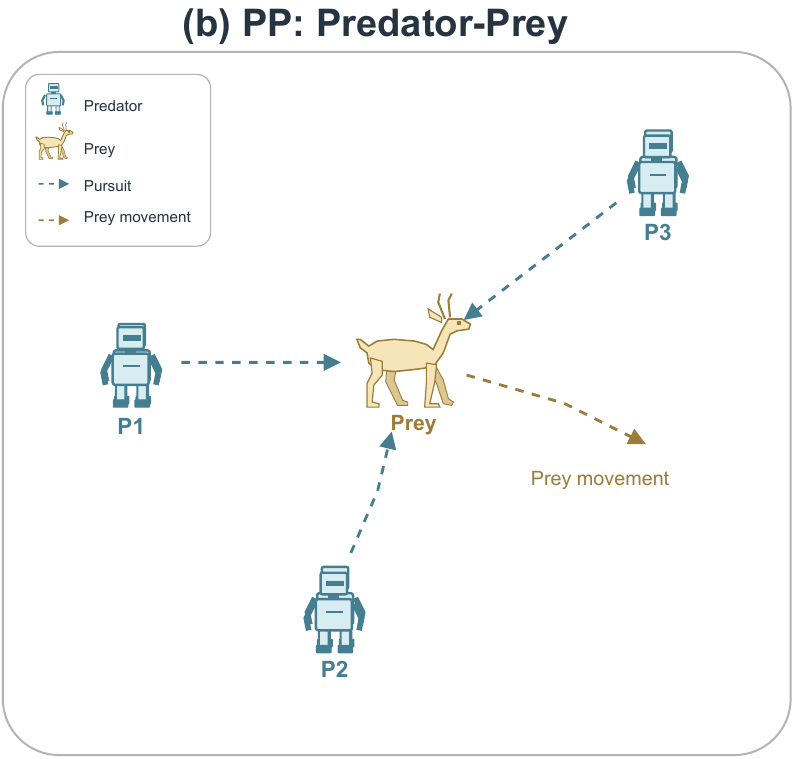}
  \end{subfigure}

  \vspace{1.5em}
  \begin{subfigure}[b]{0.80\linewidth}
    \centering
    \includegraphics[width=\linewidth]{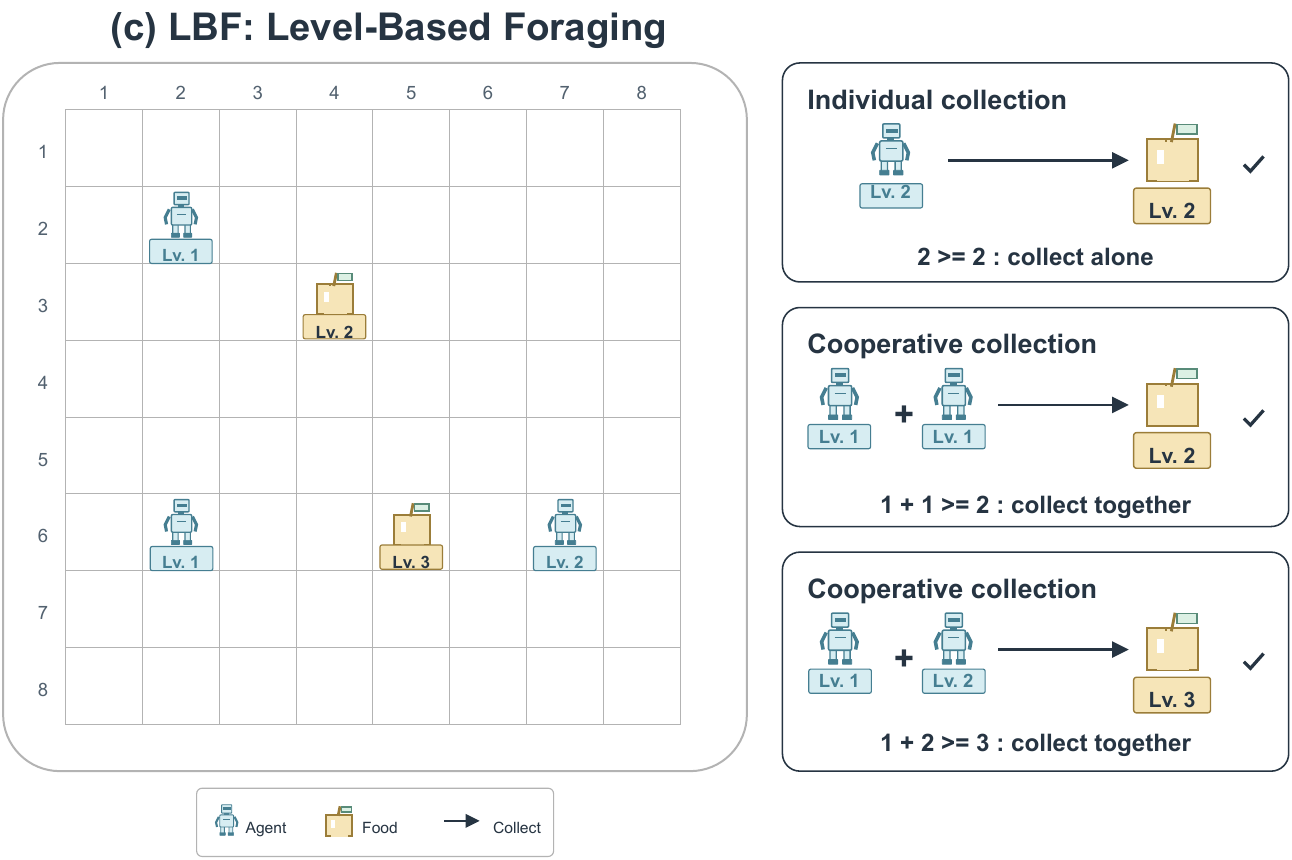}
  \end{subfigure}
  \caption{Three multi-agent environments used in our experiments. (a)~MPE: 3 agents collaboratively cover 3 landmarks. (b)~PP: 3 predators collaboratively chase a prey controlled by a pre-trained MADDPG policy. (c)~LBF: 3 agents of different levels forage 3 food items on an 8$\times$8 grid.}
  \label{fig:envs}
\end{figure}

\noindent\textbf{Multi-Agent Environments.}\quad
The experiments are conducted on three multi-agent environments.
In all environments, the agents controlled by the policy under test pursue the same task, and fairness is evaluated among them.

\begin{compactitem}
    \item
\textbf{Multi-Agent Particle Environments (MPE).} 
MPE~\cite{lowe2017multi} consists of navigation tasks, where agents control particles to cover target landmarks, characterized by dense rewards. 
We study the \textit{simple\_spread} task with 3 agents and 3 landmarks, where agents must collaboratively cover all landmarks. 

\item \textbf{Level-Based Foraging (LBF).}
LBF~\cite{papoudakis2020benchmarking} simulates a foraging scenario where agents with different levels must collect food items whose level requirements must be met by the sum of agents' levels. 
It features tasks with sparse rewards. We use a configuration with 3 agents (levels $[1, 1, 2]$) and 3 food items on a grid.

\item \textbf{Predator-Prey (PP).}
PP~\cite{lowe2017multi} models a pursuit scenario where a team of predators must collaboratively catch a prey, characterized by dense rewards. 
The \textit{simple\_tag} task with 3 predators and 1 prey is selected, where fairness is evaluated among the 3 predators.
The prey is controlled by a MADDPG~\cite{lowe2017multi} policy pre-trained in the official repository, which we directly reuse as part of the environment. 

\end{compactitem}

\noindent\textbf{Baseline Approaches.}\quad
Four baseline approaches are implemented for fairness testing in each multi-agent task.
All baselines and the proposed method share the same testing budget of 18,000 episodes per run, with 10 independent runs.

\begin{compactitem}
\item \textbf{MASTest}~\cite{ma2024enhancing}: a state-of-the-art search-based testing approach designed for multi-agent systems that maintains a corpus of test cases and evolves them through action perturbation at critical states.
It employs individual and team diversity metrics to guide the search, using an energy function that balances diversity guidance, failure feedback, and state frequency.

\item \textbf{GMT}~\cite{li2023generative}: a generative model-based testing approach that trains a diffusion model on normal initial states and generates novel test states through reverse diffusion. 
It uses novelty-based guidance to tune the diffusion model towards producing diverse states.

\item \textbf{STARLA}~\cite{zolfagharian2023search}: a state-of-the-art search-based testing approach designed for deep reinforcement learning agent.

\item \textbf{Random}: a baseline that generates test episodes by randomly sampling initial states and executing the policy with stochastic action selection. 
It serves as a reference lower bound without any guided search or learning-based generation.
\end{compactitem}

\noindent\textbf{MARL Algorithms.}\quad
For each environment, two MARL algorithms are used to train the policies under test:

\begin{compactitem}
\item \textbf{QMIX}~\cite{rashid2020monotonic}: a value-based MARL algorithm with a monotonic mixing network that decomposes the joint Q-value into individual agent Q-values, enabling centralized training with decentralized execution. 

\item \textbf{IQL}~\cite{tan1993multi}: a value-based MARL algorithm where each agent independently learns its own Q-function without any mixing network, serving as a decentralized baseline without credit assignment. 
\end{compactitem}

\noindent\textbf{Evaluation Metrics.}\quad
Two metrics are adopted to measure the testing effectiveness of each method.
\begin{compactitem}
\item \textbf{Failure count.} The number of fairness failures found by the method, where a fairness failure is an episode whose JFI is not higher than the failure threshold, which is set to 0.80 for MPE, 0.67 for LBF, and 0.80 for PP.
These thresholds are calibrated on the JFI distribution of randomly executed episodes of the trained policies, using the same threshold for both policies. 
Each threshold is chosen as the level below which an episode is clearly unfair yet not overly frequent. 
LBF 0.67 corresponds to the discrete level where one agent receives no reward while the other two share it.
\item \textbf{Failure coverage.} The diversity of the detected failures. It is computed by discretizing the joint state space into a grid and taking the fraction of the cells touched by at least one failure episode:
\begin{equation}
\mathrm{Cov}=\frac{1}{N}\Bigl|\bigcup_{e\in\mathcal{F}}\mathcal{C}(e)\Bigr|,
\label{eq:coverage}
\end{equation}
where $\mathcal{F}$ is the set of detected failure episodes and $\mathcal{C}(e)$ is the set of grid cells visited by episode $e$.
\end{compactitem}

\noindent\textbf{Hardware Platform.}\quad
All methods are implemented in PyTorch.
All experiments are launched on a server equipped with an NVIDIA vGPU of 32 GB memory, an Intel Xeon Platinum 8260 CPU, 62 GB RAM, running on Ubuntu 20.04 OS.

\subsection{Experimental Results}
\label{subsec:results}

\subsubsection{\large\emph{RQ1. \rqone}}\mbox{}\par\vspace{6pt}

In this research question, \method's effectiveness in finding more fairness failures than GMT, MASTest, STARLA, and Random Testing is investigated under the same testing budget, measured as the number of executed episodes.
The number of detected fairness failures and the state coverage of those failures are reported, and statistical significance is assessed with the non-parametric Mann--Whitney U test and the Vargha--Delaney $A_{12}$ effect size.

\begin{figure}[htbp]
  \centering
  \includegraphics[width=\linewidth]{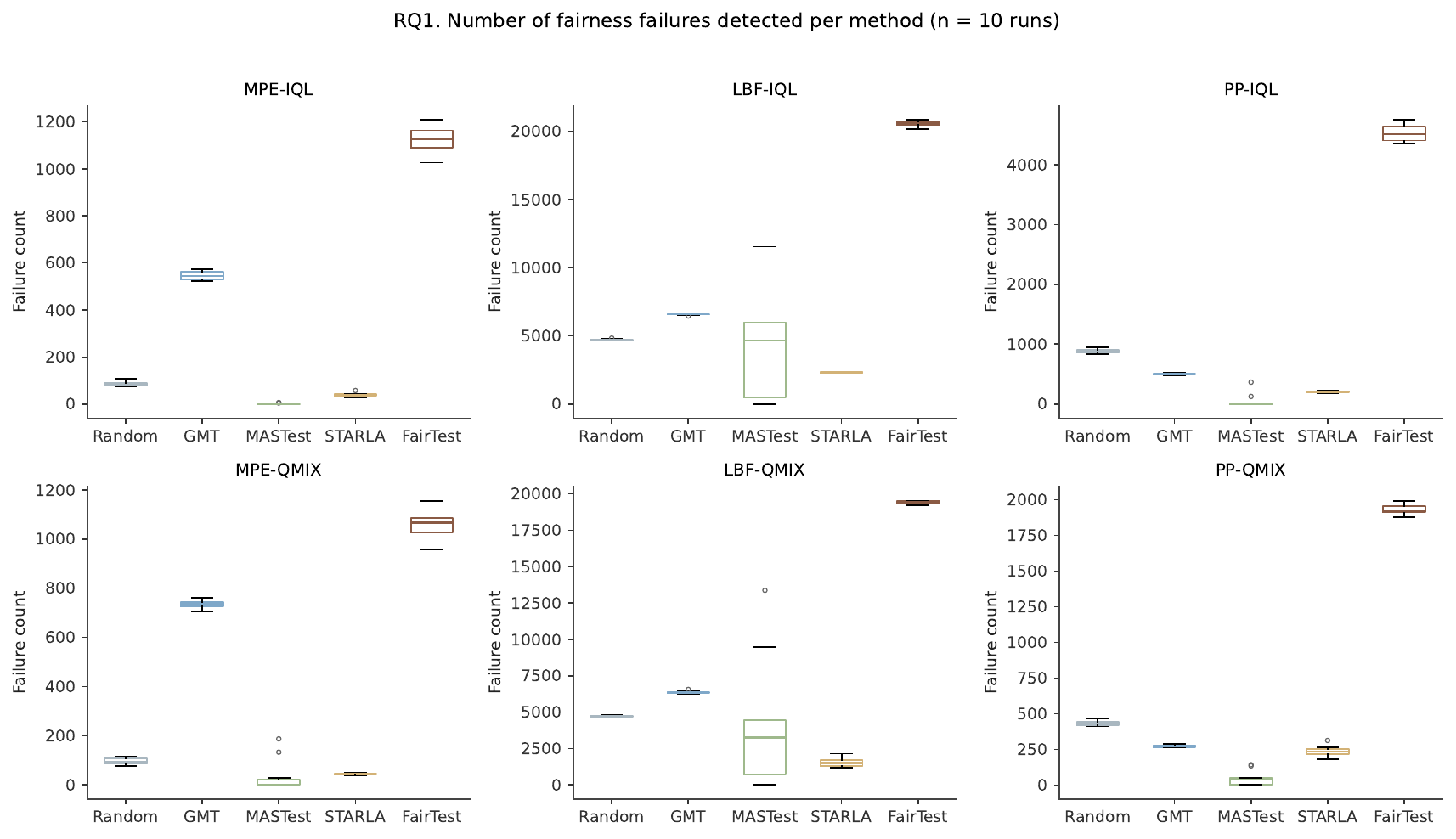}
  \vspace{-20pt}
  \captionsetup{labelformat=rq1}
  \caption{Number of fairness failures detected by each method in the six environment-algorithm configurations (10 independent runs). 
  }
  \label{fig:rq1}
\end{figure}

\noindent\textbf{Failure count.}\quad
Figure~\ref{fig:rq1} and Table~\ref{tab:rq1} report the number of fairness failures found by each method in the six environment-algorithm configurations. 
\method detects more failures than every baseline in all six configurations, and every pairwise comparison is highly significant, with $p<0.001$ and $A_{12}=1.000$, a large effect (Table~\ref{tab:rq1_stats}).
On MPE with IQL, \method reports 1{,}126.5 failures on average, against 84.3 for Random Testing and 546.4 for GMT. 
For MPE with QMIX, the counts are 1{,}055.1, 96.2, and 736.0 for \method, Random Testing, and GMT, respectively. 
Across LBF, \method yields 20{,}592.0 and 19{,}402.6 failures on average for IQL and QMIX, compared with 6{,}575.6 and 6{,}369.3 for GMT. 
On PP, it uncovers 4{,}523.4 and 1{,}930.5 failures on average for IQL and QMIX, respectively. 
MASTest behaves unstably on these fairness oracles.
From a distribution perspective, the failure count of \method is not only the highest but also the most stable across runs. 
Its coefficient of variation stays below 5.5\% in every configuration, about 5\% on MPE and below 3\% on LBF and PP, \eg, 1{,}126.5~$\pm$~57.2 on MPE-IQL and 1{,}930.5~$\pm$~34.3 on PP-QMIX. 
GMT is likewise stable, with a coefficient of variation between 0.9\% and 3.5\%, although its failure count is far lower. 
MASTest, by contrast, shows a skewed distribution with a standard deviation as large as $\pm$4{,}331 on LBF.

\begin{table}[!t]
  \centering
  \captionsetup{labelformat=rq1}
  \caption{Number of fairness failures (mean~$\pm$~std) detected by each method over 10 runs.
  \vspace{-10pt}
  }
  \label{tab:rq1}
  \resizebox{0.95\textwidth}{!}{%
  \begin{tabular}{lccccc}
    \toprule
    Env-Algorithm & Random & GMT & MASTest & STARLA & \textbf{\method}\\
    \midrule
    MPE-IQL & $84.3\pm10.4$ & $546.4\pm19.3$ & $1.0\pm2.2$ & $38.5\pm8.3$ & $\mathbf{1126.5\pm57.2}$\\
    MPE-QMIX & $96.2\pm12.9$ & $736.0\pm17.1$ & $34.8\pm67.8$ & $44.3\pm4.2$ & $\mathbf{1055.1\pm56.3}$\\
    LBF-IQL & $4701.6\pm57.0$ & $6575.6\pm59.1$ & $4322.6\pm3920.1$ & $2304.0\pm53.7$ & $\mathbf{20592.0\pm189.8}$\\
    LBF-QMIX & $4707.2\pm55.2$ & $6369.3\pm91.8$ & $4040.2\pm4331.1$ & $1555.3\pm331.3$ & $\mathbf{19402.6\pm101.4}$\\
    PP-IQL & $883.4\pm39.7$ & $498.6\pm14.3$ & $49.5\pm117.2$ & $202.4\pm18.9$ & $\mathbf{4523.4\pm134.9}$\\
    PP-QMIX & $433.4\pm18.0$ & $272.9\pm9.2$ & $47.0\pm52.0$ & $238.3\pm35.7$ & $\mathbf{1930.5\pm34.3}$\\
    \bottomrule
  \end{tabular}}
\end{table}

\begin{table}[!ht]
  \centering
  \captionsetup{labelformat=rq1}
  \caption{Statistical comparison between \method and the best baseline of each environment, determined by the highest average failure count. Two-sided Mann--Whitney U $p$-values and Vargha--Delaney $A_{12}$ effect sizes are reported for the failure count over 10 runs, separately for the IQL and QMIX.}
  \vspace{-10pt}
  \label{tab:rq1_stats}
  \small
  \setlength{\tabcolsep}{6pt}
  \begin{tabular}{lcccc}
    \toprule
    \method vs. & IQL $p$ & IQL $A_{12}$ & QMIX $p$ & QMIX $A_{12}$\\
    \midrule
    \multicolumn{5}{l}{\textbf{MPE}}\\
    Random & $0.0002^{***}$ & $1.000^{L}$ & $0.0002^{***}$ & $1.000^{L}$\\
    GMT & $0.0002^{***}$ & $1.000^{L}$ & $0.0002^{***}$ & $1.000^{L}$\\
    MASTest & $0.0001^{***}$ & $1.000^{L}$ & $0.0001^{***}$ & $1.000^{L}$\\
    STARLA & $0.0002^{***}$ & $1.000^{L}$ & $0.0002^{***}$ & $1.000^{L}$\\
    \midrule
    \multicolumn{5}{l}{\textbf{LBF}}\\
    Random & $0.0002^{***}$ & $1.000^{L}$ & $0.0002^{***}$ & $1.000^{L}$\\
    GMT & $0.0002^{***}$ & $1.000^{L}$ & $0.0002^{***}$ & $1.000^{L}$\\
    MASTest & $0.0002^{***}$ & $1.000^{L}$ & $0.0002^{***}$ & $1.000^{L}$\\
    STARLA & $0.0002^{***}$ & $1.000^{L}$ & $0.0002^{***}$ & $1.000^{L}$\\
    \midrule
    \multicolumn{5}{l}{\textbf{Predator-Prey}}\\
    Random & $0.0002^{***}$ & $1.000^{L}$ & $0.0002^{***}$ & $1.000^{L}$\\
    GMT & $0.0002^{***}$ & $1.000^{L}$ & $0.0002^{***}$ & $1.000^{L}$\\
    MASTest & $0.0001^{***}$ & $1.000^{L}$ & $0.0002^{***}$ & $1.000^{L}$\\
    STARLA & $0.0002^{***}$ & $1.000^{L}$ & $0.0002^{***}$ & $1.000^{L}$\\
    \bottomrule
    \multicolumn{5}{l}{\footnotesize{$^{*}p<0.05$; $^{**}p<0.01$; $^{***}p<0.001$; $^{L}$large effect size ($A_{12}\ge0.71$).}}\\
  \end{tabular}
\end{table}

\begin{table}[!ht]
  \centering
  \captionsetup{labelformat=rq1}
  \caption{The coverage of each method.}
  \vspace{-10pt}
  \label{tab:rq1cov}
  \resizebox{0.7\textwidth}{!}{%
  \begin{tabular}{lccccc}
    \toprule
    Env-Algorithm & Random & GMT & MASTest & STARLA & \textbf{\method}\\
    \midrule
    MPE-IQL & 0.113 & 0.386 & 0.002 & 0.055 & \textbf{0.391}\\
    MPE-QMIX & 0.131 & \textbf{0.531} & 0.009 & 0.067 & 0.441\\
    LBF-IQL & 0.888 & \textbf{0.991} & 0.064 & 0.779 & 0.985\\
    LBF-QMIX & 0.861 & \textbf{0.986} & 0.065 & 0.651 & 0.967\\
    PP-IQL & 0.570 & 0.416 & 0.006 & 0.193 & \textbf{0.889}\\
    PP-QMIX & 0.454 & 0.338 & 0.007 & 0.307 & \textbf{0.821}\\
    \bottomrule
  \end{tabular}
  }
\end{table}

\noindent\textbf{Failure coverage.}\quad
Table~\ref{tab:rq1cov} summarizes the coverage. 
\method achieves higher coverage than Random, MASTest, and STARLA in every configuration, and higher than GMT on PP, 0.889 for IQL and 0.821 for QMIX compared with 0.416 and 0.338 for GMT. 
In three configurations, however, GMT covers more cells than \method despite finding far fewer failures. On MPE-QMIX, the coverage of GMT is 0.531, and that of \method is 0.441. 
On LBF, 0.991 and 0.986 of the cells are covered by GMT for IQL and QMIX, whereas \method covers 0.985 and 0.967.
We also observe that in some cases GMT achieves a higher coverage.
For instance, on LBF, Pareto-based survivor prioritization confines the \method population to the most-unfair episodes, bounding exploration by the parent distribution. 
GMT's initial states are decoded from Gaussian noise with visited states penalized, so its fewer failures are scattered and cover more cells per failure. 
Moreover, there are also differences between the two algorithms, \ie, QMIX and IQL.
Centralized training disperses QMIX's unfair behaviors across the state space.
Independent training keeps IQL's behaviors concentrated and simple. 
Accordingly, fewer failures are detected under QMIX than under IQL in every environment, with 1{,}055.1 against 1{,}126.5 on MPE.

\begin{tcolorbox}[size=title, colback=white, breakable]
{\textbf{Answer to RQ1:}
Under the same testing budget, \method finds the most fairness failures in all six configurations, and the advantage over every baseline is statistically significant. 
Its failures also spread more widely than those of Random, MASTest, and STARLA. 
In the three configurations where GMT covers more cells, its failure count is far lower, since its exploration is scattered rather than targeted at unfair regions, and its wider coverage does not indicate higher effectiveness.
}
\end{tcolorbox}

\subsubsection{\large\emph{RQ2. \rqtwo}}\mbox{}\par\vspace{6pt}

\par\vspace{2pt}This research question investigates how the testing-budget ratio, the fraction of the candidate pool retained by the test prioritization, influences the number of fairness failures found.

\textbf{Experimental Settings. } 
\method starts each search round by generating 7{,}500 candidate episodes online through random executions of the trained agents. 
The candidate pool of 7,500 is chosen as five times the population size. 
Generating a pool larger than the retained population and filtering it down follows the MOSA-based search~\cite{panichella2015reformulating,deb2002fast}, where the excess candidates give the Pareto-front filtering a real choice and preserve the diversity of the survivors, with the crowding distance spreading the selected candidates over the front only when the pool is populated densely enough. 
A pool close to the population size would force the filtering to keep almost every candidate, and the retained population would lose its diversity.
Each candidate is executed and evaluated on three fitness functions, namely the fairness JFI $f_1$, the predicted fairness $f_2$, and the decision uncertainty $f_3$. 
The Pareto-front filtering then retains a subset of a fixed size based only on $f_2$ and $f_3$. 
It first pre-selects the top candidates by uncertainty, then applies non-dominated sorting over $f_2$ and $f_3$, and finally completes or prunes the set with the crowding distance to the target size. 
Only the retained population enters the subsequent search, so the testing budget is directly determined by the population size. The budget ratio \textit{select\_ratio} is scanned across 5\%, 10\%, 15\%, and 20\%, with retained populations of 375, 750, 1{,}125, and 1{,}500 out of the 7{,}500 candidates, where 20\% is the setting used by the proposed method.

\begin{figure}[htbp]
  \centering
  \includegraphics[width=\linewidth]{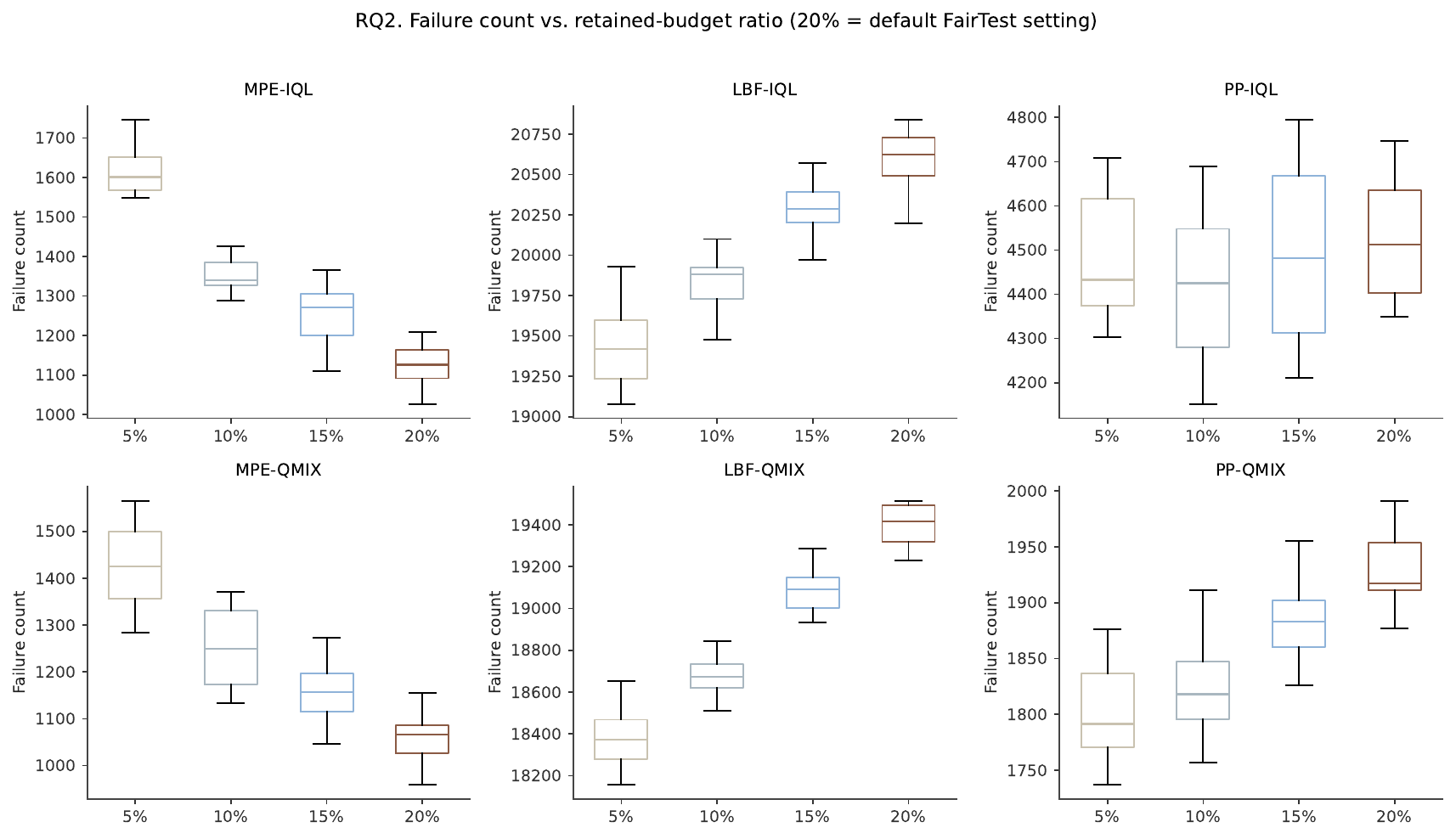}
  \vspace{-23pt}
  \captionsetup{labelformat=rq2}
  \caption{Failure count under different retained-budget ratios (5\%, 10\%, 15\%, 20\%) for the six environment-algorithm configurations.}
  \label{fig:rq2}
\end{figure}

\begin{table}[htbp]
  \centering
  \captionsetup{labelformat=rq2}
  \caption{Failure count under different retained-budget ratios (5\%, 10\%, 15\%, 20\%) for the six environment-algorithm configurations.}
  \vspace{-10pt}
  \label{tab:rq2}
  \footnotesize
  \setlength{\tabcolsep}{4pt}
  \begin{tabular}{lcccccc}
    \toprule
    Ratio & MPE-IQL & MPE-QMIX & LBF-IQL & LBF-QMIX & PP-IQL & PP-QMIX\\
    \midrule
    5\% & $\mathbf{1615.9\pm63.0}$ & $\mathbf{1429.5\pm96.8}$ & $19443.3\pm262.6$ & $18373.1\pm152.0$ & $4477.9\pm150.7$ & $1798.7\pm46.2$\\
    10\% & $1352.5\pm45.5$ & $1252.8\pm89.2$ & $19820.9\pm199.1$ & $18675.0\pm105.6$ & $4426.1\pm182.4$ & $1827.4\pm47.3$\\
    15\% & $1252.6\pm77.2$ & $1158.7\pm72.6$ & $20285.5\pm195.5$ & $19087.2\pm113.3$ & $4487.1\pm209.3$ & $1882.6\pm41.1$\\
    20\% & $1126.5\pm57.2$ & $1055.1\pm56.3$ & $\mathbf{20592.0\pm189.8}$ & $\mathbf{19402.6\pm101.4}$ & $\mathbf{4523.4\pm134.9}$ & $\mathbf{1930.5\pm34.3}$\\
    \bottomrule
    \\
  \end{tabular}
\end{table}

Figure~\ref{fig:rq2} and Table~\ref{tab:rq2} present the failure count of each ratio.
The effect of the ratio differs across environments. 
On MPE, the lower the ratio, the more failures are found. 
Shrinking the ratio from 20\% to 5\% raises the failure count from 1{,}126.5 to 1{,}615.9 for IQL and from 1{,}055.1 to 1{,}429.5 for QMIX. 
On LBF the trend is reversed, and the higher the ratio, the more failures are found, whereas PP stays essentially flat across the four ratios. 
On LBF, moving from 5\% to 20\% grows the failure count from 19{,}443.3 for IQL and 18{,}373.1 for QMIX to 20{,}592.0 and 19{,}402.6. 
On PP, the same shift lifts the counts from 4{,}477.9 for IQL and 1{,}798.7 for QMIX to 4{,}523.4 and 1{,}930.5. 
These results indicate that a stricter prioritization concentrates the budget on failure-dense regions on MPE. 
A larger retained population enables more thorough exploration on LBF and PP.

Two further observations follow. 
First, the trend is consistent between the two algorithms within each environment: over the full range of ratios, IQL and QMIX change in the same direction in every environment, which indicates that the effect of the ratio is governed by the geometry of the environment's unfair states rather than by the algorithm itself. 
Second, the sensitivity to the ratio differs sharply across environments. 
On MPE, moving from 20\% to 5\% changes the failure count by 43\% for IQL and 35\% for QMIX. 
On LBF and PP, by contrast, the total change over the whole range is at most 6\% and 7\%, respectively.
Recalling RQ1, where the default 20\% setting already outperforms all baselines, \method is robust to this parameter on LBF and PP, and on MPE its efficiency can be pushed even further by shrinking the retained population.

\begin{tcolorbox}[size=title, colback=white, breakable]
{\textbf{Answer to RQ2:}
The impact of the budget ratio on failure detection is environment-specific. 
On MPE, a smaller ratio increases the number of detected failures, and on LBF and PP, a larger ratio does the same.
In practice, it is recommended to use the default 20\% setting.
}
\end{tcolorbox}

\subsubsection{\large\emph{RQ3. \rqthree}}\mbox{}\par\vspace{6pt}

In this research question, the contribution of the three fairness-aware designs of \method is quantified, namely the fairness features used by the machine learning fairness predictor, the fitness functions that guide the search, and the Pareto-based prioritization.
All experiments are conducted on Predator-Prey under the same testing budget, and every configuration is evaluated over 10 independent runs.
RQ3-a verifies the usefulness of the fairness features in the machine learning fairness predictor with a 5-fold cross-validated AUC-ROC.
RQ3-b examines how much guidance each fitness function provides to the search, by running the search with one fitness function kept and the others removed.
RQ3-c evaluates the gain brought by the Pareto-based prioritization.

\noindent\textbf{RQ3-a: verification of the machine learning fairness features.}\quad
This sub-question verifies whether the fairness features of Section~\ref{subsec:ml} are useful to the machine learning fairness predictor on top of the original abstract states.
The predictor behind the Pareto prioritization is a random-forest classifier trained on the abstract states together with the fairness features.
The ablation removes only the fairness features and keeps the abstract states unchanged.
The fairness features summarize the outcome of the episode with the rewards of the individual agents and the statistics of the reward distribution.
The abstract states are derived from the $Q$-values and contain no reward information.
Two feature sets are compared, the abstract states alone and the full set with the fairness features added.
The metric is the 5-fold cross-validated AUC-ROC.

\begin{figure}[htbp]
  \centering
  \includegraphics[width=0.8\linewidth]{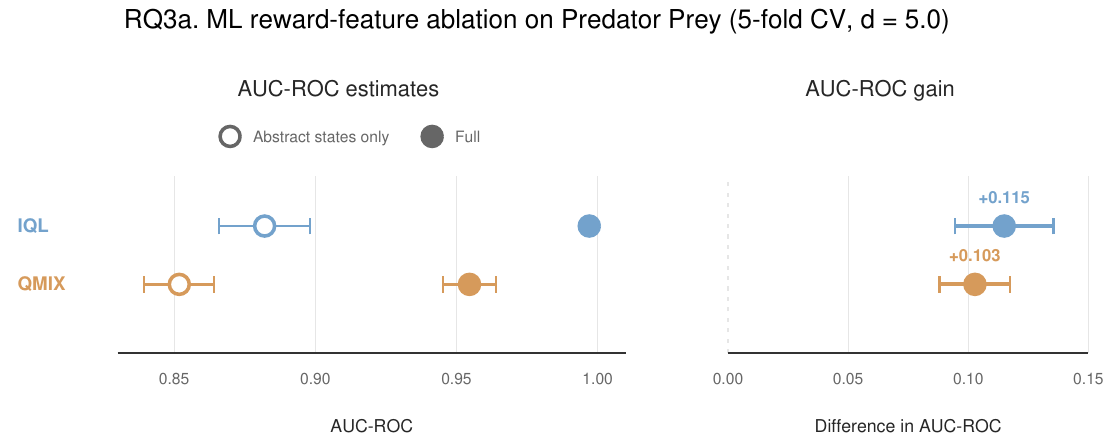}
  \vspace{-12pt}
  \captionsetup{labelformat=rq3a}
  \caption{machine learning reward-feature ablation on Predator-Prey.
  For each policy, the hollow marker is the abstract-state-only predictor and the solid marker is the predictor with the fairness features added.
  Left panel: the AUC-ROC mean with one standard deviation over the five folds.
  Right panel: the difference in AUC-ROC between the two feature sets.
  }
  \label{fig:rq3a}
\end{figure}

Adding the fairness features markedly improves the predictor, with the AUC-ROC rising from 0.882 to 0.997 for IQL and from 0.852 to 0.955 for QMIX (Figure~\ref{fig:rq3a}). 
The gain is substantial in both policies, and it is expected from the definition of the fairness index, which is computed from the cumulative rewards that these features encode.
The resulting prediction is near-perfect, so the $f_2$ guidance provided to the Pareto prioritization is reliable, and the prioritization can accurately target the candidates most likely to be unfair.

\begin{figure}[htbp]
  \centering
  \includegraphics[width=0.82\linewidth]{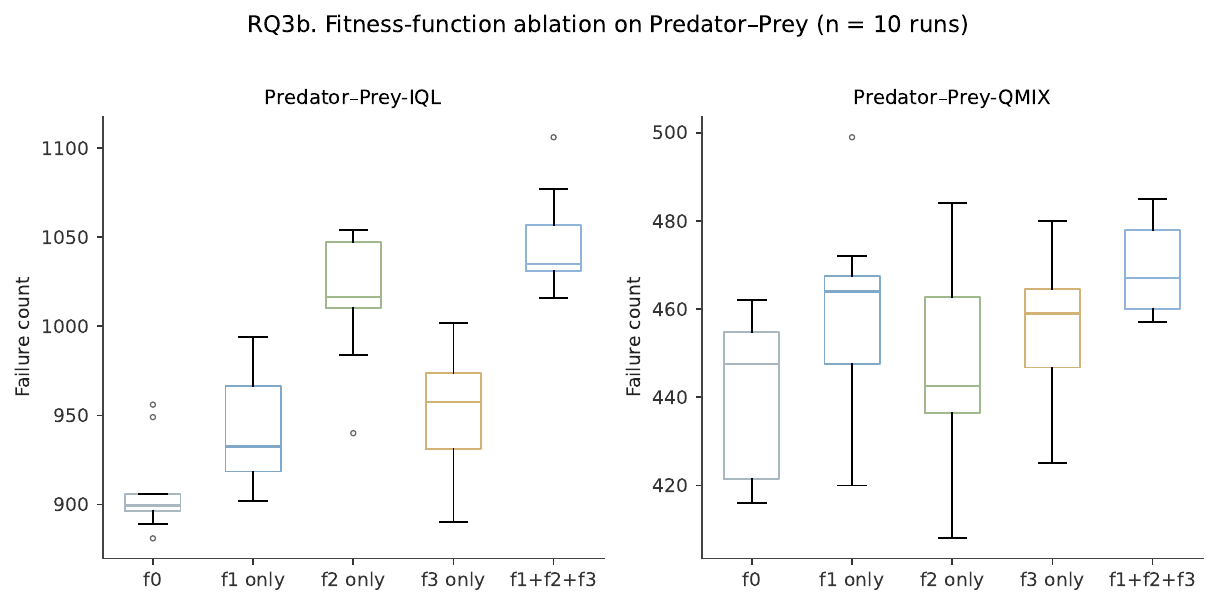}
  \vspace{-12pt}
  \captionsetup{labelformat=rq3b}
  \caption{Fitness-function ablation on Predator-Prey: failure count of the five configurations for the IQL and QMIX policies (10 runs).}
  \label{fig:rq3b}
\end{figure}

\noindent\textbf{RQ3-b: fitness function ablation.}\quad
This sub-question examines the guidance provided by each fitness function during the search.
The search of \method is driven by three fitness functions: $f_1$ is the JFI of an episode, $f_2$ is the probability of being fair predicted by the machine learning fairness predictor, and $f_3$ is the DeepGini-based decision uncertainty.
Five configurations are compared, and all of them run the same plain search framework in which no Pareto-front-based filtering is applied to the candidate pool. In each configuration, the ablated fitness functions are removed from the search and provide no guidance:
\begin{compactitem}
    \item \textbf{f0:} no fitness function is kept, and the search receives no guidance.
    \item \textbf{f1 only:} only $f_1$ is kept.
    \item \textbf{f2 only:} only $f_2$ is kept.
    \item \textbf{f3 only:} only $f_3$ is kept.
    \item \textbf{f1+f2+f3:} all three fitness functions are kept, which is the complete fitness setting of \method.
\end{compactitem}
In all five configurations, a fairness failure is always determined by the real JFI of an episode, and this common criterion keeps the comparison across the configurations fair.

Figure~\ref{fig:rq3b} reports the failure count of the five configurations, and every configuration with an enabled fitness function detects more failures than the unguided f0 on both policies.
On IQL, the count reaches 942.3 with f1, 1{,}018.2 with f2, and 951.4 with f3, while f0 stays at 907.9.
The predicted fairness f2 is the strongest single signal, and its gain of about 12\% is consistent with RQ3-a, where the fairness features make the prediction near-perfect, so the episodes preferred along f2 are the ones that are actually unfair.
On QMIX, f1, f2, and f3 detect 459.2, 446.3, and 454.5 failures, respectively, all above the 441.1 of f0, and the direct fairness signal f1 contributes the most among the single objectives.
Unfair episodes are rarer on this policy, and the exploration driven by uncertainty reaches them more easily than a direct fairness signal.
The full setting f1+f2+f3 records 1{,}046.1 failures on IQL and 469.1 on QMIX, slightly above the best single function of each policy, 1{,}018.2 and 459.2.
The three fitness functions all provide useful guidance, and every configuration with an enabled fitness function outperforms the unguided one.

\begin{figure}[htbp]
  \centering
  \includegraphics[width=1.0\linewidth]{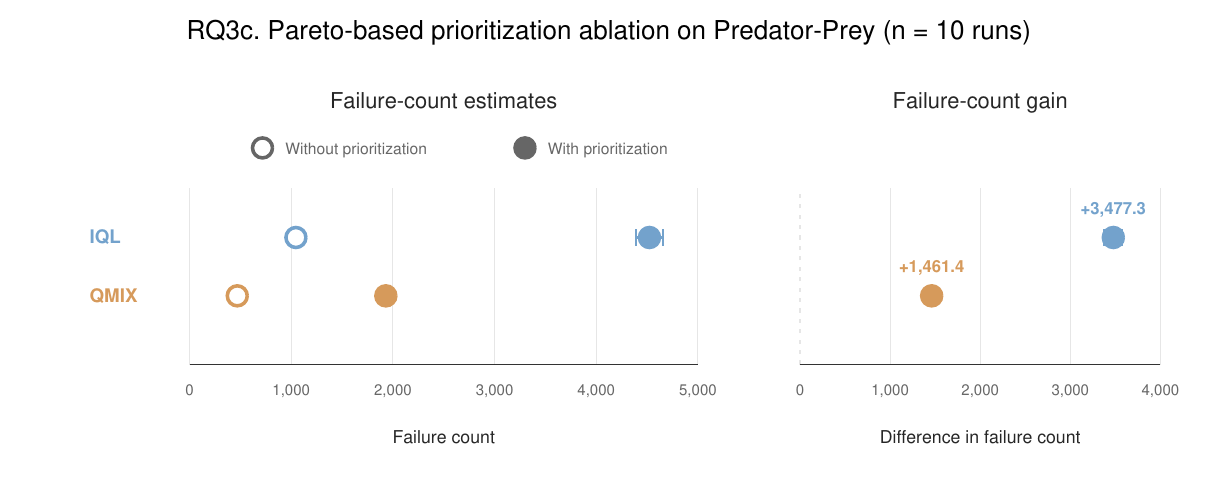}
  \vspace{-30pt}
  \captionsetup{labelformat=rq3c}
  \caption{Pareto-based prioritization ablation on Predator-Prey (n = 10 runs).
Left panel: the failure count of the search without (hollow) and with (solid) the Pareto-based prioritization, shown as the mean with one standard deviation over the runs for each policy.
Right panel: the gain in failure count brought by the prioritization.}
  \label{fig:rq3c}
\end{figure}

\noindent\textbf{RQ3-c: Pareto-based prioritization.}\quad
This sub-question isolates the contribution of the Pareto-based prioritization by comparing the same search with and without it.
The two configurations share the same fitness functions and the same testing budget, and they differ only in whether the candidates of each generation are screened by the Pareto front of $f_2$ and $f_3$.

Figure~\ref{fig:rq3c} reports the comparison. 
With the prioritization, the failure count reaches 4,523.4 on IQL and 1,930.5 on QMIX, about four times the 1,046.1 and 469.1 of the run without it, and the coverage rises to 0.889 and 0.821, up from the 0.495 and 0.379. 
Both configurations execute the same number of episodes, and the gain comes from which episodes are executed rather than from how many. 
In the run without the prioritization, the executed episodes are drawn without any screening, so most of them are predicted to be fair and only a small fraction of them turns into a failure. 
The prioritized run spends the same budget on the candidates that are most likely to be unfair: the candidates are filtered by the Pareto front of $f_2$ and $f_3$, the survivors score low on both objectives at the same time, and the crowding distance keeps them spread out over the score space. 
Each executed episode then has a much higher chance of becoming a failure, which raises both the failure count and the coverage. 
Together with RQ3-b, the comparison shows that the fitness functions guide the search, and the Pareto-based prioritization turns that guidance into a concentrated use of the budget, which is the main source of the effectiveness of \method.

\begin{table}[H]
  \centering
  \captionsetup{labelformat=rq3c}
  \caption{Failure count (mean~$\pm$~std) and coverage of the search with and without the Pareto-based prioritization on Predator-Prey.}
  \vspace{-10pt}
\label{tab:rq3c}
  \small
  \setlength{\tabcolsep}{6pt}
  \begin{tabular}{lcccc}
    \toprule
    Configuration & IQL failure count & QMIX failure count & IQL cov. & QMIX cov.\\
    \midrule
    Without prioritization & $1{,}046.1\pm27.8$ & $469.1\pm10.7$ & 0.495 & 0.379\\
    \textbf{With prioritization (\method)} & $\mathbf{4{,}523.4\pm134.9}$ & $\mathbf{1{,}930.5\pm34.3}$ & \textbf{0.889} & \textbf{0.821}\\
    \bottomrule
  \end{tabular}
\end{table}

\begin{tcolorbox}[size=title, colback=white, breakable]
{\textbf{Answer to RQ3:}
The fairness features encode the reward structure behind the fairness index, and they make the predictor trustworthy for guiding the search (RQ3-a).
Each fitness function alone improves on the unguided configuration, and the most useful single signal depends on the policy, with the predicted fairness on IQL and the direct fairness signal on QMIX (RQ3-b).
The prioritization focuses the execution budget on the episodes with the greatest chance of being unfair, which amplifies the detected failures and spreads them wider (RQ3-c).
}
\end{tcolorbox}

\section{Threats to Validity}
\label{sec:threats}

\noindent\textbf{Internal threats} concern whether the reported failures really happen in the environment. 
Candidates are chosen based on the fitness scores, and a candidate picked as unfair may turn out to be fair when it is executed, which would make the reported failures wrong. 
This is addressed by execution, since a failure is counted only after its episode is executed, with the JFI computed from the actual rewards of that trajectory. 
The verification covers both operators, as mutation runs the episode again from the changed state, while crossover offspring are executed after the search. 
The manually chosen threshold is another threat, for a value that is too high or too low would skew the comparison. 
Each threshold is calibrated from the JFI values of random executions of the trained policies. 

\noindent\textbf{External threats} come from the limited scope of the experiments, and the findings may not transfer to other systems.
Training the policies and executing thousands of episodes for every method and run is computationally expensive, and this cost confines the scope to three environments and two MARL algorithms.
The chosen settings mitigate the threat.
The three environments are widely used benchmarks~\cite{lowe2017multi,papoudakis2020benchmarking} and cover three task structures, navigation, foraging, and pursuit.
The two policies belong to two families of value-based MARL algorithms, with and without credit assignment.
The search operators and the fitness functions are defined on the recorded transitions of any policy, so the approach does not depend on a specific environment.

\noindent\textbf{Construct threats} arise from whether the metrics and the predictor measure the right quantities. 
The JFI measures how evenly the rewards are divided among the agents, and it says nothing about the size of the rewards, with two allocations such as $(3, 1, 2)$ and $(6, 2, 4)$ receiving the same value even though their sizes differ. 
The size is outside the fairness question studied here, which only compares the division, and no necessary information is lost.
The failure count raises a second concern.
A single count cannot distinguish many similar failures from many different ones, and the coverage complements it by measuring how the failures spread over the state space.
The predictor raises a third concern.
A wrong prediction may waste the budget on fair episodes, but the reported failures are judged by the measured JFI of the executed episodes, and the predictor only decides which episodes are run. 
Its accuracy changes the search efficiency, while the validity of the reported failures stays the same, with the accuracy reported in the experiment evaluation.

\section{Related Work}
\label{sec:related_work}

\subsection{Fairness in Multi-Agent Reinforcement Learning}
\label{subsec:mar-fairness}

Fairness in MARL has been studied as a metric or a training objective, but rarely as a testing target. 
Jain's Fairness Index (JFI)~\cite{jain1984quantitative} is one of the standard metric, with $J=1$ indicating perfect equity. 
The Gini coefficient and the min-max ratio are also commonly used. 
On the training side, AdaFair-MARL~\cite{ekpo2025adafair} enforces JFI as an explicit fairness constraint via primal-dual updates.
EcoFair-CH-MARL~\cite{alqithami2026ecofair} uses the Gini coefficient and min-max ratio with an $O(\sqrt{T})$ regret bound. 
AFP~\cite{wei2026integrated} integrates altruistic and fairness preferences into the utility. 
Fairness applications also appear in domains including stock trading~\cite{bao2019fairness}. 
For credit assignment, COMA~\cite{foerster2018counterfactual} provides a counterfactual baseline, while SHAQ~\cite{wang2020shapley} approximates the Shapley value~\cite{shapley1953value} and Nucleolus~\cite{li2025nucleolus} applies the nucleolus solution concept from cooperative game theory, both achieving fair credit assignment across coalitions. 
All these works optimize fairness at training time, yet none validates whether a trained policy actually produces fair executions at deployment: a policy may achieve high average reward while systematically starving some agents of reward in certain episodes, failures that surface only at the trajectory level and cannot be ruled out by training-time fairness constraints. 
This is the gap addressed by our method \method. 
It treats episodes with low JFI as fairness failures and, within a search-based testing framework, borrows the multi-objective search of MOSA~\cite{panichella2015reformulating} with NSGA-II-style Pareto-front selection~\cite{deb2002fast}: each generation, candidates are scored on the predicted fairness and Q-value decision uncertainty, and the survivors are kept for the next round, extending MARL fairness from training-time optimization to pre-deployment verification.

\subsection{Testing Multi-Agent Reinforcement Learning}
\label{subsec:mar-testing}

MARL algorithms (MADDPG~\cite{lowe2017multi}, QMIX~\cite{rashid2020monotonic}, MAPPO~\cite{yu2022surprising}, COMA~\cite{foerster2018counterfactual}), benchmarked on EPyMARL~\cite{papoudakis2020benchmarking} and SMAC~\cite{samvelyan2019starcraft}, pose greater testing challenges because the joint action space grows exponentially with the number of agents. 
Existing MAS testing research proceeds along three lines. 
Adversarial attacks are the most studied evaluation approach: an attacker injects perturbations into agents' observations, actions, or communication, or trains adversarial policies, to induce errors in the target policy. 
Huang \etal{}~\cite{huang2017adversarial} first introduce adversarial examples to neural network policies, Zhang \etal{}~\cite{zhang2021robust} impose state perturbations with a learned optimal adversary, and Ilahi \etal{}~\cite{ilahi2021challenges} survey attack types such as observation perturbation, adversarial policies, and communication attacks, together with defenses. 
AdapAM~\cite{chen2026adversarial} by Chen \etal{} adaptively selects the victim agent at each step under a strict black-box setting and generates stealthy perturbations via a proxy model. 
Their companion work EMAI~\cite{chen2025understanding} measures the importance of individual agents in a team through counterfactual reasoning, supporting policy understanding, attack, and patching. 
Robustness testing evaluates MAS from a failure-detection perspective: Zhou and Liu's RTCA~\cite{zhou2023robustness} selects critical agents via differential evolution, advises worst-case joint actions, and perturbs their observations. Guo \etal{}~\cite{guo2022towards} conduct a systematic robustness evaluation of cooperative MARL. 
Beyond attacks and robustness, diversity-guided search-based testing is another important line, and MASTest~\cite{ma2024enhancing} by Ma \etal{} is, to the best of our knowledge, the only testing approach designed specifically for MAS. 
MASTest models the behavior of a team as a graph: the trajectories of individual agents are embedded with node2vec~\cite{grover2016node2vec}, and the Wasserstein Weisfeiler-Lehman graph kernel~\cite{togninalli2019wasserstein} measures the distance between such team-behavior graphs, so the search can explore behaviors that are diverse from the perspective of the whole team. 
It further identifies critical states that are error-prone in historical executions and adaptively exploits them to trigger more failures. 
This explicit modeling of inter-agent structure is what distinguishes MASTest from policy-level methods such as MDPFuzz and GMT, which can be migrated to MAS but regard the system as a whole and do not capture how agents interact. 
The test oracles of all the above works target functional failures or robustness degradation. 
None treats fairness as a testing objective. Unlike these methods, we adopt fairness as the test oracle and follow MASTest's idea of diversity guided search to find low JFI episodes in MARL, which is an objective untouched by existing MAS testing work.

\subsection{Testing of Single-Agent Deep Reinforcement Learning}
\label{subsec:single-agent-testing}

Deep Reinforcement Learning (DRL) approximates policies with deep neural networks and the resulting agents have been widely deployed in autonomous driving, robotics, and games. 
Testing DRL agents is therefore essential to the reliability of such systems~\cite{mnih2015human,sutton1998reinforcement,11570074,XIE2026112763,bellman1966dynamic}. 
Testing research for single-agent DRL has drawn attention from both the software engineering and the artificial intelligence communities, covering correctness, robustness, and safety monitoring. 
Search-based testing, \eg, STARLA~\cite{zolfagharian2023search}, uses genetic algorithms guided by machine-learning failure-prediction models to target failing episodes, and INDAGO~\cite{biagiola2024testing} trains such a model on the configurations in which the agent failed during training, which detects 50\% more failures than the state of the art. 
Fuzz testing (MDPFuzz~\cite{pang2022mdpfuzz}, SeqDivFuzz~\cite{wang2023fuzzing}, and coverage-guided fuzzing~\cite{wan2024coverage}) mutates initial states under heuristics such as state-coverage density and sequence diversity to reveal low-reward executions, as systematically evaluated by Kang \etal{}~\cite{kang2026evaluating}. 
GMT~\cite{li2023generative} is a generative approach that synthesizes test cases with diffusion models guided by novelty. 
The prediction of whether an execution will fail and the choice of which executions to run are the two ingredients that the present approach shares with the methods below. 
SelfOracle~\cite{stocco2020misbehaviour} predicts the failure before it happens by monitoring the confidence of a driving model at run time, and Foresee~\cite{naziri2026misbehavior} forecasts near misses that guide local fuzzing. 
PRT~\cite{lin2026failure} chooses the task regions where failures are more likely, which halves the cost of finding a failure compared with random testing. 
Other lines include BehAVExplor~\cite{cheng2023behavexplor} for behavior-diversity testing, MOSAT~\cite{tian2022mosat} for multi-objective safety testing, SMARLA~\cite{zolfagharian2024smarla} for runtime monitoring, DRLMutation~\cite{li2025drlmutation}, $\mu$PRL~\cite{thomas2025muprl}, RLMutation~\cite{tambon2023mutation} and MDPMorph~\cite{li2025mdpmorph} for mutation testing, coverage criteria for test adequacy~\cite{shi2024multigranularity}, reinforcement learning in the role of the test generator~\cite{romdhana2022deep,giamattei2024reinforcement} and adversarial generation~\cite{pasini2025cross}, and plasticity testing~\cite{biagiola2022plasticity}.
These methods differ in focus but generally rely on specific environments or hand-crafted heuristics, and their failure definitions are mostly limited to reward crashes and safety violations of a single agent, making it hard to cover complex objectives including multi-agent interaction and fairness. 
The systematic review by Sunba \etal{}~\cite{sunba2025testing} likewise identifies ``specialized testing for fairness'' as a key gap. 
In contrast, our method targets fairness testing of the policies of multi-agent systems: it combines search-based testing and machine-learning guidance, prioritizes candidate episodes via Pareto-front filtering~\cite{deb2002fast,hu2024test}, and detects fairness failures with low JFI within a limited budget.

\section{Conclusion}
\label{sec:conclusion}

This paper has presented \method, which verifies the fairness of MARL policies through search-based testing before deployment.
The approach detects unfair executions through the search guidance and the test prioritization.
For the guidance, the search scores every candidate with the fairness measured on the runs already performed, the fairness predicted from abstract states and fairness features, and the decision uncertainty of the policy. 
Crossover and mutation then produce further candidates by varying the executions already run.
For the prioritization, the runs are directed by the predicted fairness and the decision uncertainty to the candidates where failures are expected.
A failure is reported when its execution confirms that the fairness lies below the threshold. 
The failures found by \method outnumber those of every baseline in all configurations, and the differences are statistically significant with a large effect size.
The ablations confirm the value of the fairness features and of each fitness function, and the Pareto-based prioritization delivers the largest part of the gain.
The budget-ratio study shows that the best ratio depends on the environment, and the default setting stays competitive in all environments.
Several directions remain open for future work.
For example, the failures that \method found can be fed back into retraining as a way of fairness enhancement for the policy.
Moreover, root cause analysis involves localizing the sub-episodes that cause a fairness failure, which can identify the responsible agent and the corresponding action.

\bibliographystyle{ACM-Reference-Format} \bibliography{sample-base}
\end{document}